\documentclass[a4paper,11pt]{article}
\usepackage{jcappub} % for details on the use of the package, please see the JINST-author-manual
\usepackage{lineno}
\usepackage{orcidlink}
\usepackage{cleveref}
\usepackage{multirow}
\usepackage{makecell}
\usepackage{array}

\usepackage{booktabs} 
\usepackage{xcolor}
\usepackage{comment}
\usepackage[normalem]{ulem}
\usepackage{amssymb}

\arxivnumber{1234.56789} % Only if you have one
\title{\boldmath Magnetized protoneutron stars: Structure and stability}

\author[a]{Harsh Chandrakar \orcidlink{0009-0001-7994-4789}}
\affiliation[a]{Department of Physics, BITS-Pilani K. K. Birla Goa Campus, Goa 403726, India}

\author[b,c]{Adamu Issifu \orcidlink{0000-0002-2843-835X}} 
\affiliation[b]{Departamento de F\'isica, Instituto Tecnol\'ogico de Aeron\'autica, DCTA, 12228-900, S\~ao Jos\'e dos Campos, SP, Brazil} 
\affiliation[c]{Laborat\'orio de Computa\c c\~ao Cient\'ifica Avan\c cada e Modelamento (Lab-CCAM), Brazil}

\author[d]{Prashant Thakur~\orcidlink{0000-0001-5930-7179}}
\affiliation[d]{Department of Physics, Yonsei University, Seoul, 03722, South Korea}

\author[a]{T. K. Jha \orcidlink{0000-0002-9334-240X}}
\author[a]{Aravind Taridalu \orcidlink{0000-0003-4189-6176}}
\emailAdd{prashant@yonsei.ac.kr}

\abstract{ We investigate the evolution of magnetized protoneutron stars (PNSs) across four schematic stages: neutrino-trapped, deleptonization, neutrino-transparent, and the final cold, catalyzed neutron star (NS). Using a quasi-static approximation on the Kelvin–Helmholtz timescale, we model strongly magnetized configurations ($B \sim 10^{17}\,\text{G}$) using axisymmetric XNS 4.0 with Equations of State (EoS) derived from relativistic mean-field theory calibrated by the DDME2 parameterization. We analyze the gravitational mass, equatorial radius, shape deformation, magnetic flux, and magnetic-to-binding energy ratio as functions of thermodynamic and compositional changes. Our results show that increasing entropy per baryon and decreasing lepton fraction raise core temperatures, which in turn enhance magnetic deformation, flux confinement, and the magnetic-to-binding energy ratio. Magnetic field dissipation is most efficient during the deleptonization and neutrino-transparent stages, a process that ultimately determines the observable field strength of the mature NS. This work presents the first general-relativistic characterization of how PNS thermal and compositional evolution reshapes magnetic field deformation and energetics across poloidal, toroidal, and mixed configurations at fixed baryonic mass.}

\begin{document}
\maketitle
\flushbottom

\section{Introduction}
\label{sec:intro}

The ultra-compact remnants of massive stellar explosions, known as NSs, provide special cosmic laboratories for the study of matter at the extremes of gravity and density \cite{Burrows:1986me}. These intriguing objects do not form instantly; rather, they emerge from a brief, violent, and thermally dynamic phase, during which the nascent object is referred to as a PNS~\cite{Pons:1998mm}. Understanding the PNS phase is therefore critical to understanding the final properties of mature NSs~\cite{Lattimer_2004}. A PNS is born in the immediate aftermath of a core-collapse supernova or a binary NS merger. It is characterized by extreme conditions, including very high temperatures that can reach up to 50~MeV in the core and a high entropy per baryon of $s_B \approx 1{-}2$ \cite{ Prakash:1996xs}. In this hot and dense environment, neutrinos become trapped on dynamical timescales, leading to a large, trapped lepton fraction and a finite neutrino chemical potential \cite{Bethe,Burrows:1986me, Mazurek1975}.  Over a period of approximately 10--20 seconds, the PNS undergoes significant evolution as it cools and deleptonizes, primarily through the diffusion and emission of these trapped neutrinos~\cite{1995A&A...296..145K, PhysRevLett.104.251101, Roberts_2017}. Despite being a highly dynamic process, a few seconds after the initial core bounce, the PNS evolution enters the Kelvin-Helmholtz phase~\cite{Pons:1998mm, Fischer:2009af}, where it can be reasonably approximated as a sequence of quasi-equilibrium configurations. This evolutionary path, from a hot, lepton-rich object to a cold, catalyzed NS, is governed by a complex interplay of the strong, weak, electromagnetic, and gravitational forces, and has a profound impact on the star's structure and composition.

Ultrastrong interior magnetic fields, on the order of \(10^{18}\,\text{G}\), can significantly influence compact stars in two primary ways. First, they interact with the constituent particles inside the star, thereby modifying the EoS~\cite{Chatterjee:2021wsr,Chatterjee:2014qsa,Rather:2022bmm,Scurto:2024xxo, Malik:2018blm}. Second, such intense fields alter the energy-momentum tensor~\cite{Bocquet:1995je,Cardall:2000bs,Kiuchi:2008ss,Pili:2014npa,Ioka:2003nh}, leading to the breaking of spherical symmetry and consequently affecting the star's external structure.
PNSs are expected to harbor powerful magnetic fields, a feature inherited and dramatically amplified from their progenitors \cite{Spruit_2008, Ferrario_2015}. While observations of magnetars suggest surface fields reaching $10^{15}{-}10^{16}$~G \cite{Mereghetti_2008, Kaspi}, theoretical arguments based on the virial theorem predict that internal fields could be orders of magnitude stronger, potentially reaching $\sim10^{18}$~G \cite{Ferrer,1991ApJ...383..745L,1989ApJ...342..958F, Cardall_2001}. The amplification from a progenitor star's weaker field is thought to occur through two primary mechanisms during the star's violent birth: the conservation of magnetic flux during gravitational collapse \cite{1964ApJ...140.1309W} and dynamo processes driven by the PNS's convection and differential rotation\cite{1992ApJ...392L...9D}. Additionally, magnetorotational instabilities (MRIs) may play a role in rapidly amplifying seed magnetic fields \cite{Akiyama_2003, Obergaulinger_2009, Sawai_2013, M_sta_2015}. Given the star's high conductivity, these intense fields are considered to be ``frozen-in'' during the short evolutionary timescale of a PNS \cite{Chamel_2008,RevModPhys.80.1455}. Such enormous magnetic fields are not passive components; they fundamentally alter the PNS's structure and composition \cite{Lattimer_2001, Oertel:2016bki}. The Lorentz force exerted by the field induces a significant deformation of the star, with the geometry of the distortion depending on the field's configuration \cite{1953ApJ...118..116C, Pili:2014npa}. A purely poloidal field makes the star oblate (flattened at the poles), while a purely toroidal field results in a prolate (elongated) shape. Crucially, the extent of this deformation is strongly linked to the star's thermal state; a hot PNS is more susceptible to distortion and will display a greater ellipticity than a cold NS for the same magnetic field strength. This deformation has direct observational consequences, as a rotating, deformed star is a potential source of continuous gravitational waves \cite{2011CQGra..28k4014G, Cutler_2002,2007Ap&SS.308..119D, Mastrano, bonazzola1996gravitationalwavesneutronstars}.

Beyond modifying the static structure, magnetic fields dynamically affect key physical processes that govern the evolution of a PNS. They influence neutrino transport, alter the core's chemical composition, and affect the development of hydrodynamical instabilities. In particular, strong magnetic fields impact neutrino transport by introducing anisotropies in neutrino-matter interaction cross-sections~\cite{Luo:2024qmq,PhysRevC.107.065804}. The likelihood of neutrino scattering or absorption depends on the angle between the neutrino's trajectory and the magnetic field lines. Such fields can also suppress efficient cooling mechanisms like the Direct Urca process, leading to longer thermal relaxation times. The magnetic pressure reduces the central baryon density, which may delay or inhibit the formation of exotic components like hyperons or deconfined quark matter~\cite{PhysRevLett.79.2176, PhysRevD.54.1306}. Additionally, a strong poloidal field leads to an oblate stellar shape, introducing asymmetries in temperature and composition that can result in anisotropic neutrino emission from the surface~\cite{P_rez_Azor_n_2006}.

To study magnetic fields in PNSs, one must use the framework of General Relativistic Magnetohydrodynamics (GRMHD)~\cite{Sur:2021awe, Cheong:2024stz, Das:2022oyn}, which couples fluid dynamics with Maxwell's and Einstein's equations. However, evolving the full dynamics of spacetime and matter is computationally expensive. A common and well-justified simplification replaces this dynamic evolution with a sequence of static, quasi-equilibrium configurations\cite{Pons:1998mm,Camelio:2017nka}. This is valid under strong magnetic fields (\(B \gtrsim 10^{14} \, \text{G}\)), where the Alfvén timescale is much shorter than the thermal evolution timescale (\(\sim 10\text{--}20\) s), allowing the field to equilibrate rapidly. The extended Conformally Flat Condition (XCFC) further simplifies the metric by assuming spatial conformal flatness. Codes like \texttt{XNS 4.0}~\cite{Bucciantini:2014zca, refId1} use this approach to efficiently compute magnetized equilibrium structures in general relativity.

In this study, we have modeled non-rotating PNSs to systematically explore the effects of strong magnetic fields on their global properties across the four key stages of PNS evolution. The main objective is to understand how the internal magnetic field configuration and the evolving EoS interact to influence the structure and deformation of the PNS. By constructing non-rotating models, we are able to isolate magnetic effects without the complicating influence of rotation. This allows us to attribute any observed stellar deformation directly to magnetic pressure and tension, thereby enabling a clearer analysis of magnetic field-induced changes. To ensure a self-consistent comparison across all four evolutionary stages, we fix the baryonic mass of our models at 1.92~$M_{\odot}$. This isolates the effects of the evolving EoS, guaranteeing that all variations in the star's gravitational mass, radius, and deformation are a direct result of changes in its internal thermodynamics and composition. This method eliminates the confounding variables that would arise from comparing different stellar progenitors, thus providing a clean framework for our analysis. Our current approach, therefore, serves as a necessary and well-established step in developing a comprehensive understanding of magnetised PNSs.

Additionally, dividing the evolution of newly formed PNSs into four separate regimes, neutrino trapping, deleptonization, neutrino-transparent, and a cold, catalyzed configuration—serves two purposes. First, it allows us to verify our calculated global quantities against well-established models for the initial and final stages, which enhances the trustworthiness of our results. Second, it enables a concentrated examination of the predominantly under-researched function of internal magnetic field geometry during the intermediate stages, especially during the deleptonization and neutrino-transparent phases.
This phase-resolved method also helps us separate the effects of changing thermodynamic conditions, like changes in entropy, lepton content, and neutrino opacity, from purely magnetic effects. By keeping the baryonic mass the same at all stages, we make sure that any changes in structure we see, like changes in radius, deformation, and magnetic energy content, are mostly due to the way the EoS and magnetic field configuration change together, not because of differences in total mass or composition.

The structure of this paper is as follows: Section~\ref{NS1} discusses the microphysical inputs relevant to the evolution of PNS. Section~\ref{XNS} outlines the formalism and implementation of magnetic field effects using the \texttt{XNS 4.0} code. The results are presented and analyzed in Section~\ref{results}, followed by the main conclusions and key insights in Section~\ref{conclusion}.

\section{Microphysics}\label{NS1}

{To study the interaction between the hadrons (protons and neutrons), we used the quantum hadrodynamics (QHD) formalism \cite{Serot:1997xg}, in which the exchange of massive mesons simulates the strong force between them. Considering the massive mesons $\sigma$ (scalar-isoscalar), $\omega$ (vector-isoscalar), and $\rho$ (vector-isovector) mesons, the Lagrangian density that describes the system can be expressed as: 
\begin{align}
 \mathcal{L}_{\rm H}{}&=  \sum_{N=p,n} \bar \psi_N \Big[  i \gamma^\mu\partial_\mu - \gamma^0  \big(g_{\omega N} \omega_0  +  g_{\rho N} I_{3} \rho_{03}  \big)\nonumber\\
 &- \Big( m_N- g_{\sigma N} \sigma_0 \Big)  \Big] \psi_N, \label{h}\\
\mathcal{L}_{\rm m}&= - \frac{1}{2} m_\sigma^2 \sigma_0^2  +\frac{1}{2} m_\omega^2 \omega_0^2   +\frac{1}{2} m_\rho^2 \rho_{03}^2,\label{m}\\
 \mathcal{L}_{\rm l}& = \sum_l\Bar{\psi}_l\left(i\gamma^\mu\partial_\mu-m_l\right)\psi_l\label{l},
\end{align}
Here, $\psi_N$ denotes the baryonic Dirac field for nucleons, with $m_N = 938\,\text{MeV}$ being the nucleon mass. The meson masses are represented by $m_i$ ($i = \sigma, \omega, \rho$), and $g_{iN}$ are the corresponding nucleon-meson coupling constants. $I_3 = \pm 1/2$ denotes the isospin projection of the nucleons. The subscript `0' of the meson fields denotes their mean-field equivalent. The term $\psi_l$ refers to the free lepton field, introduced into the stellar matter to ensure charge neutrality, as NSs are physically observable objects. The leptonic sector includes all relevant leptons present in supernova remnants, namely electrons, electron neutrinos, and muons \cite{Issifu:2023qyi, Issifu:2024fuw, Issifu:2023qoo, Prakash:1996xs, Pons:1998mm}. The model is calibrated using the DDME2 parameterization~\cite{ddme2}, represented by: 
\begin{equation}
    g_{i N} (n_B) = g_{iN} (n_0)a_i  \frac{1+b_i (\eta + d_i)^2}{1 +c_i (\eta + d_i)^2},
\end{equation}
and 
\begin{equation}
    g_{\rho N} (n_B) = g_{\rho N} (n_0) \exp\left[ - a_\rho \big( \eta -1 \big) \right].
\end{equation}

\begin{table}[]
\caption {DDME2 parameters.}
\begin{center}
\begin{tabular}{ |c| c| c| c| c| c| c| }
\hline
 meson($i$) & $m_i(\text{MeV})$ & $a_i$ & $b_i$ & $c_i$ & $d_i$ & $g_{i N} (n_0)$\\
 \hline
 $\sigma$ & 550.1238 & 1.3881 & 1.0943 & 1.7057 & 0.4421 & 10.5396 \\  
 $\omega$ & 783 & 1.3892 & 0.9240 & 1.4062 & 0.4775 & 13.0189  \\
 $\rho$ & 763 & 0.5647 & $\cdots$ & $\cdots$ & $\cdots$ & 7.3672 \\
 \hline
\end{tabular}
\label{tab:T}
\end{center}
\end{table}

Where $\eta = n_B / n_0$, with $n_0 = 0.152\,\text{fm}^{-3}$ representing the nuclear saturation density for this parameterization, and $n_B$ being the baryon density. The model parameters ($a_i$, $b_i$, $c_i$,\, and $d_i$) are fitted to reproduce experimental bulk nuclear matter properties near $n_0$. Further details on the binding energy, incompressibility modulus, symmetry energy, and its slope under this parameterization are given in \cite{Roca-Maza:2011alv, Reed:2021nqk}. The parameters of the model, including the corresponding coupling constants and meson masses, are presented in \cref{tab:T}.

To study the evolutionary stages of PNSs, we assume charge neutrality and $\beta$-equilibrium in the stellar matter. The corresponding EoS is determined by solving the energy-momentum conservation equations of Eqs.~(\ref{h}), (\ref{m}), and (\ref{l}), i.e. $\mathcal{L}=\mathcal{L}_{\rm H}+\mathcal{L}_{\rm m}+\mathcal{L}_{\rm l}$~\cite{Menezes:2021jmw}. Detailed procedures for calculating the EoS for PNSs within this formalism are provided in Refs.~\cite{Raduta:2020fdn, Issifu:2023qyi, Issifu:2024htq}. Additional references for cold NS matter using similar relativistic mean-field approaches with density-dependent couplings can be found in Refs.~\cite{Typel:2018cap, Roca-Maza:2011alv, Niksic:2002yp}. The EoS is evaluated for four distinct stages of NS evolution, each characterized by different thermodynamic conditions, following the treatments in Refs.~\cite{Prakash:1996xs, Pons:1998mm, Prakash:2000jr}. The first two stages correspond to neutrino-trapped matter: The hot, lepton-rich phase immediately after core collapse, when the NS is born, and the deleptonization phase, which occurs shortly after core bounce as the trapped neutrinos begin to diffuse outward. For these neutrino-trapped stages, the thermodynamic condition is governed by:
\begin{equation}\label{s2}
sT = P + \varepsilon - n_B\mu_B - \mu_{\nu_e} (n_{\nu_e} + n_e),
\end{equation}
where $s$ is the entropy density, $T$ is the temperature, $P$ is the pressure, $\varepsilon$ is the energy density, $\mu_B$ is the baryon chemical potential, $\mu_{\nu_e}$ is the electron neutrino chemical potential, $n_{\nu_e}$ is the neutrino number density, and $n_e$ is the electron number density. The final two stages correspond to neutrino-transparent matter: The early post-neutrino escape phase, where the star begins to cool, and the cold, catalyzed NS phase, representing the star's long-term equilibrium state \cite{Issifu:2024fuw, Issifu:2024htq}. For these neutrino-transparent stages, the thermodynamic condition simplifies to:
\begin{equation}\label{s1}
sT = P + \varepsilon - n_B\mu_B.
\end{equation}
Therefore, by applying the equations presented above, we can determine the EoS properties and the corresponding temperature distributions of the stellar matter at each evolutionary stage. The primary distinction between the neutrino-trapped and neutrino-transparent stages lies in the role of neutrino pressure: During the neutrino-trapped phases, neutrinos contribute significantly to the pressure, helping to support the star against gravitational collapse. In contrast, once the neutrinos have escaped, this pressure support vanishes, making the star more susceptible to gravity-driven contraction.

 Finally, in the model framework, we neglect photons and non-electron neutrinos (muon and tau). The leptonic sector in the Lagrangian \eqref{l} and the corresponding EoS include only electrons, electron neutrinos ($\nu_e$), and muons, which dominate charge neutrality and $\beta$-equilibrium during the PNS's evolution stages considered. {In the PNS regime, the high-density, post-bounce core where baryon densities reach $\rho_B \gtrsim 0.1\,\rho_0$ and temperatures are $T \sim 10$--$50~\mathrm{MeV}$.  The energy density and pressure are predominantly determined by baryons, degenerate electrons, and trapped electron neutrinos, while contributions from photons and other neutrino flavors are subdominant. This is consistent with PNS modeling assumptions, as adopted in \cite{Pons:1998mm, Prakash:1996xs, Raduta:2020fdn, Burrows:1986me, Malfatti:2019tpg, Shao:2011nu}. For example, at $T = 30$--$50~\mathrm{MeV}$ and $\rho_B / \rho_0 \approx 0.4$--$4$, the photon energy density is $\varepsilon_\gamma \approx 0.007$--$0.5~\mathrm{MeV\,fm^{-3}}$, less than 1\% of the baryonic rest-mass energy density ($\varepsilon_\mathrm{rest} \approx 60$--$600~\mathrm{MeV\,fm^{-3}}$). Including baryonic interactions and degenerate leptons further reduces the photon fraction.}

\section{Modeling PNS in a strong magnetic field using XNS 4.0}\label{XNS}

To investigate the structure of strongly magnetized PNSs across their evolutionary stages, we construct static, axisymmetric equilibrium configurations within the framework of general relativity. The numerical solutions are obtained using the open source \texttt{XNS 4.0} code \citep{Bucciantini:2014zca, Pili:2014npa}, which is based on the extended Conformally Flat Condition (XCFC) formalism \citep{Cordero-Carrion:2008grk}. For a complete description of the underlying numerical formalism, including the metric and the governing Grad-Shafranov and Bernoulli equations, we refer an interested readers to  Ref.~\cite{Pili:2014npa}, for more details.

A key input for our models is EOS for each of the four evolutionary stages, derived from the relativistic mean-field theory calibrated by the DDME2 parameterization as detailed in Section~\ref{NS1}. We investigate three magnetic field geometries by specifying the free functions, $\mathcal{M}$ and $\mathcal{I}$, used within the XNS formalism to control the field structure. Our approach for each geometry is as follows:

\subsubsection{Poloidal and Mixed Configurations}

For stellar models incorporating a poloidal magnetic field, including purely poloidal and mixed-field configurations, the equilibrium structure is governed by the magnetization function $\mathcal{M}(A_\phi)$ and the current function $\mathcal{I}(A_\phi)$, where $A_{\phi}$ is the magnetic flux function. We adopt the functional forms described in Ref.~\citep{Ciolfi_2009}, where the poloidal magnetization function is given by:
\begin{equation}
    \mathcal{M}(A_{\phi}) = k_{\text{pol}} \left( A_{\phi}+ \frac{1}{2}\xi A^2_{\phi} \right),
    \label{eq:M_func}
\end{equation}
and the toroidal current function is:
\begin{equation}
    \mathcal{I}(A_{\phi}) = \frac{a}{\zeta + 1} \Theta[A_{\phi} - A^{\text{max}}_{\phi}](A_{\phi} - A^{\text{max}}_{\phi})^{\zeta + 1}.
    \label{eq:I_func}
\end{equation}
where $k_{\text{pol}}$ is the poloidal magnetization constant, and $\xi$ governs the nonlinear contribution to the poloidal field, with $a$ the twisted torus magnetization constant, $\zeta$ the corresponding index, and $\Theta[\cdot]$ the Heaviside step function. The quantity $A^{\text{max}}_{\phi}$ denotes the maximum value of the vector potential. In our analysis, we make specific choices for these parameters:
\begin{itemize}
    \item For purely poloidal configurations, we set $a=0$ (no toroidal component) and $\xi=0$ (linear current model).
    \item For mixed-field configurations, we use a non-zero $a$ and $k_{\text{pol}}$, but maintain linearity by setting $\xi=0$ and $\zeta=0$.
\end{itemize}

\subsubsection{Purely Toroidal Configurations}

In the case of purely toroidal magnetic fields, the magnetic flux function $A_{\phi} = 0$ throughout the star, rendering the Grad-Shafranov formalism based on $A_\phi$ inapplicable. Instead, the equilibrium structure is described using functions of the scalar quantity $G = \rho h \alpha^2 \psi^4 r^2 \sin^2\theta$. The toroidal magnetic field is then expressed as $B_{\phi} = \alpha^{-1} \mathcal{I}(G)$. For the current function $\mathcal{I}(G)$, where the current function $\mathcal{I}(G)$ is defined as:
\begin{equation}
    \mathcal{I}(G) = K_m G^m,
    \label{eq:I_G_func}
\end{equation}
which leads to a corresponding magnetization function $\mathcal{M}(G)$:
\begin{equation}
    \mathcal{M}(G) = -\frac{m K_m^2}{2m - 1} G^{2m - 1}.
    \label{eq:M_G_func}
\end{equation}
Here, $K_m$ is the toroidal magnetization constant and $m$ is the magnetization index. For this work, we fix the index at $m=1$.

For the primary results presented in this work, we analyze sequences of models constructed for each magnetic geometry and for each of the four evolutionary stages at a fixed baryonic mass of $1.92\,M_\odot$ and a maximum internal magnetic field strength of $B_{\text{max}} = 5.67 \times 10^{17}\,\mathrm{G}$.

\section{Results and Discussion} \label{results}

Before presenting our results, we note that in our sequences the maximum magnetic field strength, $B_{\text{max}}$, is held fixed while the stellar radius evolves according to the thermodynamic changes at each stage of the stellar evolution sequence. Consequently, magnetic flux is not conserved. This deliberate choice isolates the structural impact of a field of given peak strength during PNS cooling and contraction, without mixing thermodynamic evolution with flux amplification from contraction.

\begin{figure}[htbp]
    \centering
    \includegraphics[width=0.8\columnwidth]{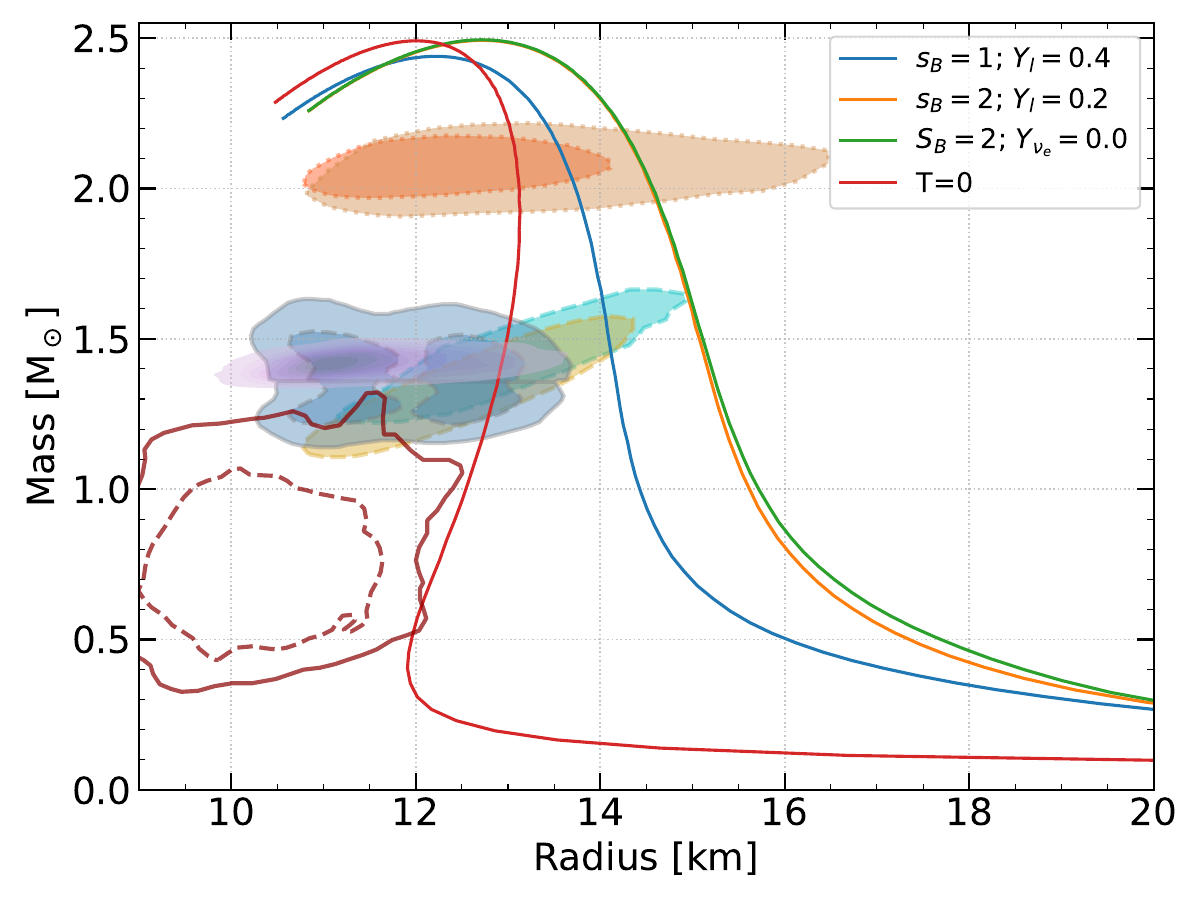}
    \caption{The plot shows the relationship between the radius and the gravitational mass, $M_{\odot}$ at different stages of PNS evolution, up to the final stage when the star becomes cold and catalyzed at $T=0$, calculated for unmagnetized case with TOV. The steel blue area indicates the constraints obtained from the binary components of GW170817, with their respective 90\% and 50\% credible intervals. Additionally, the plot includes the 1 $\sigma$ (68\%) CI for the 2D mass-radius posterior distributions of the millisecond pulsars PSR J0030 + 0451 (in cyan and yellow color) \cite{riley2019, Miller:2019cac} and PSR J0740 + 6620 (in orange and peru color)\cite{riley2021, Miller:2021qha}, based on NICER X-ray observations. Furthermore, we display the latest NICER measurements for the mass and radius of PSR J0437-4715 \cite{Choudhury:2024xbk} (lilac color). The supernova remnant HESS J1731$-$347 \cite{2022NatAs...6.1444D} is shown in red, with the outer contour representing the 90\% CL and the inner contour representing the 50\% CL.}
    \label{fig:single}
\end{figure}

In this section, we discuss how intense internal magnetic fields (on the order of $10^{17}$~G) influence the equilibrium configurations of non-rotating PNSs during their evolutionary phases. The equilibrium configurations were constructed using the \texttt{XNS 4.0} code. The quasi-equilibrium framework was considered an ideal tool for this investigation because it enables us to systematically isolate and quantify how a star's structural response is influenced by the evolving interaction between its magnetic field and thermodynamic state--particularly entropy per baryon and lepton fraction variations. This approach offers fundamental insights into these critical interactions, which would be obscured by the complexities of time-dependent instabilities found in a full dynamical simulation. Our primary objective is to comprehensively examine how strong magnetic fields impact key global stellar parameters of evolving PNSs, using the formalism presented in Ref.~\cite{Pili:2014npa}. These parameters define the macroscopic characteristics of the PNSs \cite{Prakash:1996xs, Janka:2006fh}. Specifically, we analyze how these global stellar properties depend not only on the magnitude of the internal magnetic field but also on its spatial configuration, encompassing variations in the internal distribution of electric currents and the resulting magnetic field geometry, such as purely poloidal, purely toroidal, or mixed-field arrangements. One of the key parameters we investigate is the mean deformation rate $\bar{e}$, defined via the principal moments of inertia\cite{kiuchi2008.78.044045} \begin{equation}
\bar{e} = \frac{I_{zz} - I_{xx}}{I_{zz}},
\label{eq:deformation_e}
\end{equation}
where $I_{zz}$ and $I_{xx}$ are the moments of inertia parallel and orthogonal to the symmetry axis, respectively. To capture the evolutionary progression, this analysis is conducted for four distinct stages of the star's evolution, each corresponding to a particular stage in the thermal and compositional evolution of the PNS. For consistency and comparative purposes, all stellar models are fixed at a baryonic mass of $1.92\, M_{\odot}$. The advantage of this approach is that it enables us to trace the evolution of a single star from its birth to maturity. Along the evolutionary path, both the gravitational mass and radius are expected to change due to deleptonization and the heating of stellar matter. The only quantity expected to remain conserved throughout the star's evolution is the baryon number \cite{Pons:1998mm, Nakazato:2020ogl}. Additionally, this mass was selected based on an extensive exploration of sequences of magnetized NS configurations obtained by varying the central baryon density across all four stages. The baryonic mass of $1.92\, M_{\odot}$ emerged as the optimal choice since it corresponds to the greatest average deformation and most pronounced variations in global stellar quantities across all four stages when subjected to varying magnetic field strengths. Unmagnetized reference configurations with the same baryonic mass of $1.92\, M_{\odot}$ were also constructed for each stage, facilitating a clear comparative analysis of the effects induced by magnetic fields.

{Figure~\ref{fig:single} displays the gravitational mass as a function of radius of PNSs as computed using the general Tolman–Oppenheimer–Volkoff (TOV) equations~\cite{PhysRev.55.374}, tracing their evolution from hot, neutrino-rich configurations at birth to the formation of a cold, catalyzed NS configuration several years later. The resulting mass-radius relations from this modeling are compared against observational constraints. The steel-blue shaded region represents the constraints from the binary components of GW170817, corresponding to the 90\% (outer contour) and 50\% (inner contour) confidence levels (CL). We also include the $1\sigma$ (68\%) credible intervals derived from the NICER X-ray observations of millisecond pulsars: PSR J0030+0451 (shown in cyan and yellow)~\cite{Riley:2019yda, Miller:2019cac}, and PSR J0740+6620 (in orange and peru)~\cite{Riley:2021pdl, Miller:2021qha}. In addition, the recent NICER mass-radius measurement of PSR J0437--4715~\cite{Choudhury:2024xbk} is indicated in lilac. The supernova remnant HESS J1731$-$347 \cite{doroshenko2022strangely, abramowski2011new} is shown in brown, with the outer contour representing the 90\% CL and the inner contour representing the 50\% CL. Further discussion of the thermodynamic conditions governing PNS evolution and their influence on stellar structure can be found in Refs.~\cite{Malfatti:2019tpg, Issifu:2023qyi, Pons:1998mm} and references therein.

\subsection{PURELY TOROIDAL FIELD}

The interplay between the magnetization prescription and the EoS governs the internal magnetic field structure in magnetized PNSs. For a purely toroidal configuration, constructed via the functional form of $\mathcal{M}(G)$ in Eq.~(\ref{eq:M_G_func}), the EoS dependence is clearly evident. Figure~\ref{fig:Tor_isocontous} presents isocontours of $|B|$ for equilibrium models at four evolutionary stages, each defined by a distinct EoS: (a) $s_B = 1$, $Y_l = 0.4$; (b) $s_B = 2$, $Y_l = 0.2$; (c) $s_B = 2$, $Y_{\nu_e} = 0$; and (d) $T = 0$. Stage (a) represents a suitable post-bound equilibrium stage that serves as a lepton rich initial stage for comparison \cite{Pons:1998mm, Burrows:1986me, Prakash:1996xs}. %The uniform entropy profile assumed for the initial stage (a) is a deliberate simplification, adopted to establish a clean thermodynamic baseline to gain qualitative insight for a systematic study. This idealization, which aligns with the standard approach used in foundational PNS studies \cite{Pons:1998mm, Burrows:1986me, Prakash:1996xs}, enables a controlled investigation into how global stellar properties respond to varying characteristic thermal conditions, serving as a necessary first step towards models with realistic, time-dependent profiles.

\begin{table*}[htbp]
\hspace*{-1.0cm}   % Shift table left to prevent right overflow
\setlength{\tabcolsep}{3pt}
\renewcommand{\arraystretch}{1.1}

\begin{tabular}{ccccccccccc}
\hline
Thermodynamic\\ conditions &
\begin{tabular}[c]{@{}c@{}}$\rho_c$\\$(10^{14}\,\mathrm{g/cm^3})$\end{tabular} &
\begin{tabular}[c]{@{}c@{}}$M$\\(M$_\odot$)\end{tabular} &
\begin{tabular}[c]{@{}c@{}}$r_e$\\(km)\end{tabular} &
$r_p/r_e$ &
\begin{tabular}[c]{@{}c@{}}$R_{\text{circ}}$\\(km)\end{tabular} &
\begin{tabular}[c]{@{}c@{}}$\bar{e}$\\$(10^{-1})$\end{tabular} &
\begin{tabular}[c]{@{}c@{}}$\Phi$\\$(10^{21}\,\mathrm{Wb})$\end{tabular} &
\begin{tabular}[c]{@{}c@{}}$\mathcal{H}/\mathcal{W}$\\$(10^{-1})$\end{tabular} &
\begin{tabular}[c]{@{}c@{}}$T_c$\\(MeV)\end{tabular} \\
\hline
$s_B=1,\; Y_l = 0.4$         & 5.88 & 1.72 & 16.33 & 1.12 & 18.87 & -6.03 & 13.59 & 1.72 & 18.88 \\
$s_B=2,\; Y_l = 0.2$         & 5.35 & 1.75 & 18.97 & 1.17 & 21.49 & -9.90 & 17.53 & 2.30 & 43.28 \\
$s_B=2,\; Y_{\nu_e} = 0$     & 5.40 & 1.74 & 19.02 & 1.16 & 21.56 & -9.36 & 17.25 & 2.24 & 44.30 \\
$T=0$                        & 6.46 & 1.69 & 12.17 & 1.14 & 14.69 & -5.00 & 11.24 & 1.41 & 0.0 \\
\hline
\end{tabular}
\caption{Global physical quantities of the equilibrium models with baryonic mass $m_b=1.92\,M_\odot$ and maximum magnetic field strength (purely toroidal) $B_{\max}=5.67\times10^{17}\,\mathrm{G}$. Here, $\rho_c$ is the central baryon density, $M$ the gravitational mass, $r_e$ the equatorial radius, $r_p$ the polar radius, $R_{\text{circ}}$ the circumferential radius, $\bar{e}$ the deformation parameter, $\Phi$ the magnetic flux, $\mathcal{H}$ the total magnetic energy, $\mathcal{W}$ the binding energy, and $T_c$ the core temperature.}
\label{tab:table_2}
\end{table*}

\begin{figure}[htbp]
\centering

% Row 1
\includegraphics[width=0.49\textwidth]{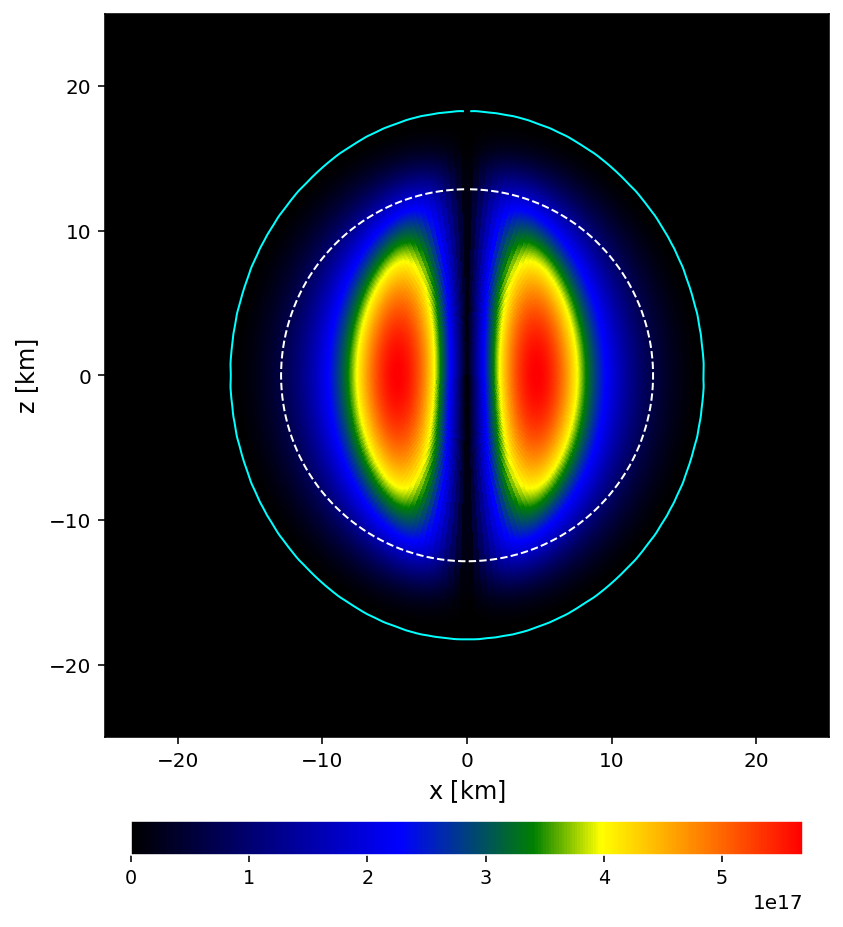}
\hfill
\includegraphics[width=0.49\textwidth]{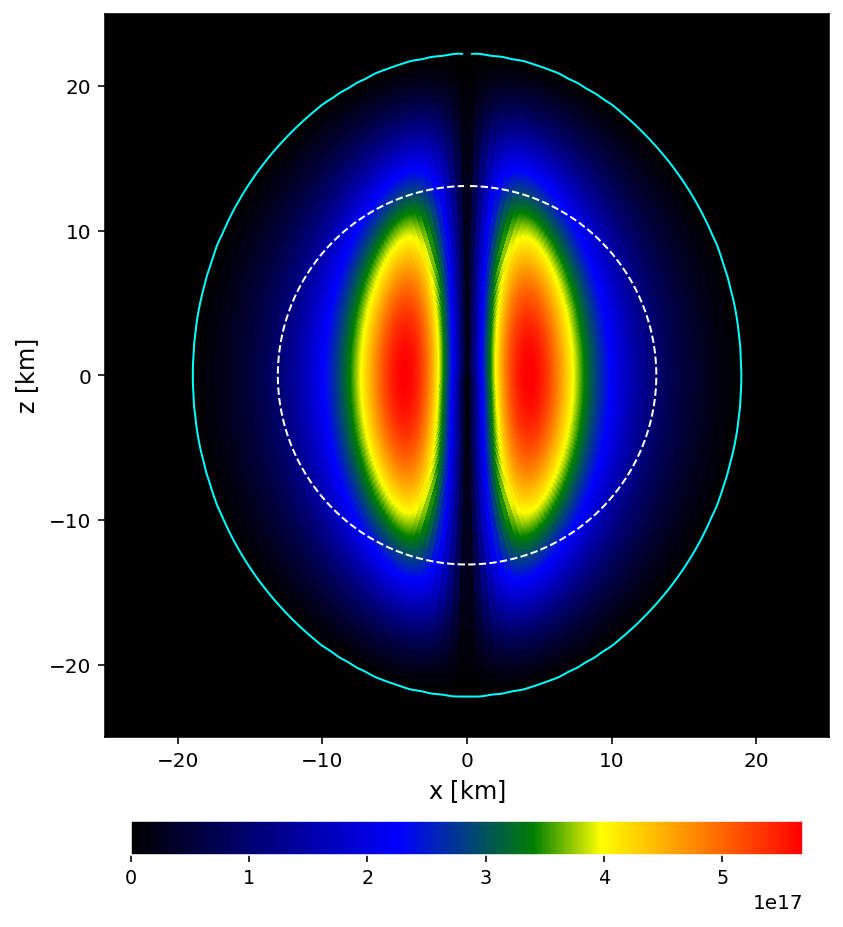}

\vspace{0.5cm}

% Row 2
\includegraphics[width=0.49\textwidth]{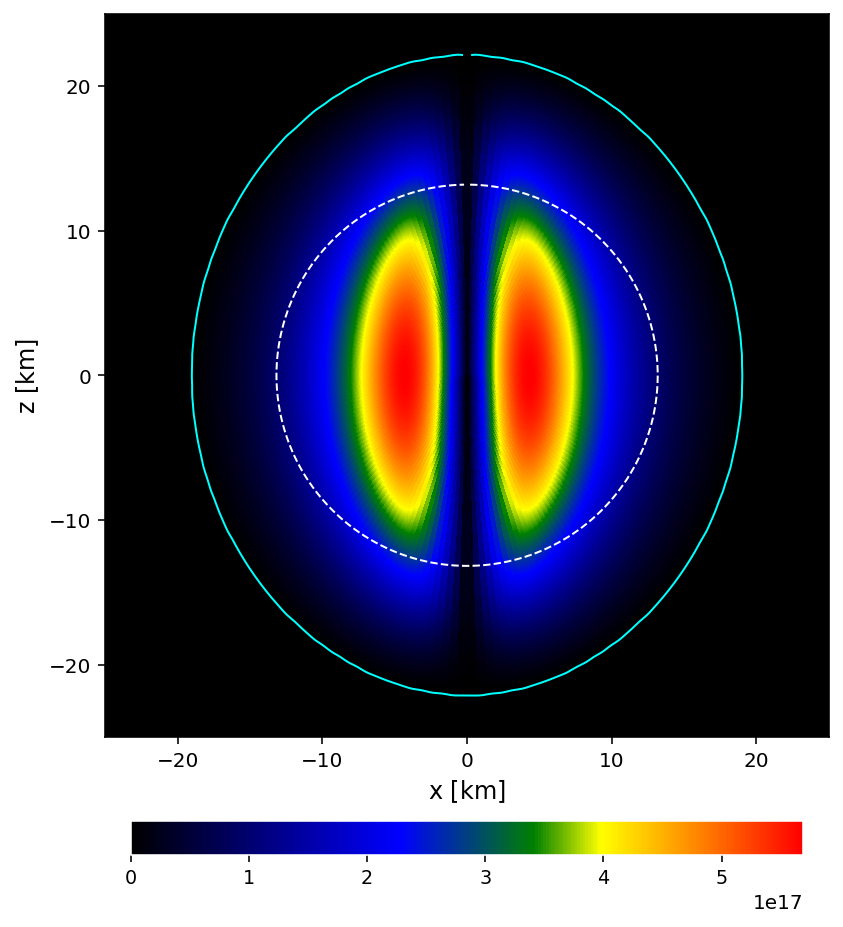}
\hfill
\includegraphics[width=0.49\textwidth]{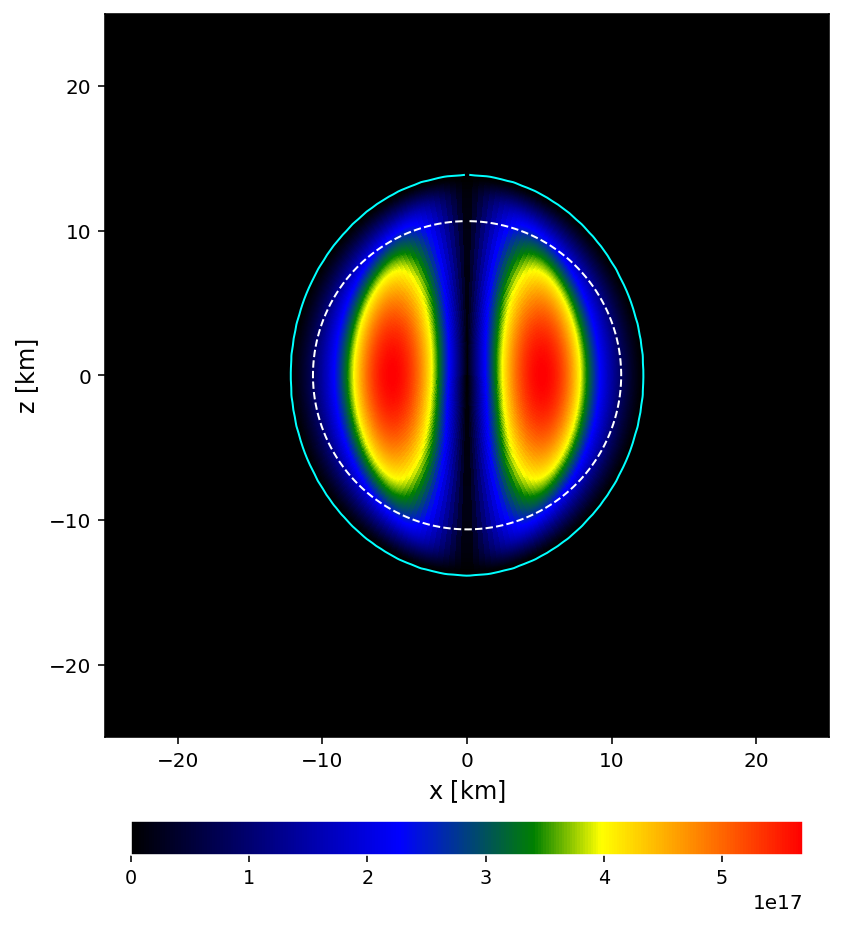}

\caption{Meridional cross-sections of PNSs with a purely toroidal magnetic field at four evolutionary stages: (a) neutrino-trapped ($s_B=1$, $Y_l=0.4$), (b) deleptonizing ($s_B=2$, $Y_l=0.2$), (c) neutrino-transparent ($s_B=2$, $Y_{\nu_e}=0$), and (d) cold and catalyzed ($T=0$). The color map shows the isocontours of magnetic field strength ($|B|$) in Gauss, while the solid cyan and dashed white lines indicate the surfaces of the magnetized and unmagnetized reference stars, respectively. All configurations share the same baryonic mass ($1.92\,M_{\odot}$) and maximum magnetic field ($B_{\text{max}} = 5.67 \times 10^{17}$\,G).}
\label{fig:Tor_isocontous}
\end{figure}

 The four stages (a)–(d) represent schematic, quasi-static equilibrium configurations at characteristic entropy and lepton fraction values, rather than a direct temporal evolution. Stage (a) represents, relatively cold and lepton-rich baseline configuration ($s_B = 1$, $Y_l = 0.4$), providing a well-defined reference point for anchoring the subsequent evolutionary stages. Stage (b) represents the hot, lepton-rich post-bounce PNS formed after shock heating, whose larger gravitational mass reflects (as shown in tables.~\ref{tab:table_2} to \ref{tab:table_mix2}) the incorporation of high-entropy material. Stage (c) corresponds to the neutrino-transparent phase, where the star, having lost its neutrino pressure support, continues to cool and contract. This evolution continues to stage (d), the final cold, catalyzed NS formation.

Figure~\ref{fig:Tor_isocontous} presents meridional cross-sections, where $x$ and $z$ are spatial coordinates in kilometers, illustrating magnetic field surfaces and the distribution of magnetic field strength. The color scale indicates the magnetic field strength in Gauss, with warmer colors (red, yellow) representing higher field strengths and cooler colors (green, blue) denoting lower strengths. The solid cyan line marks the boundary of the magnetized star, while the dashed white line shows the surface of the corresponding unmagnetized star with the same baryonic mass for reference. Comparing the white inner contours of the unmagnetized stars along the evolutionary path, we observe that the size of the star increases from (a) to (c) and then shrinks again at (d). In particular, there is a significant increase in size at stages (b) and (c), which corresponds to the deleptonization phase when neutrinos are escaping from the stellar core. During these stages, the core temperature of the PNSs is higher compared to the other stages, as shown in \cref{tab:table_2}. The same table indicates that the intermediate stages are associated with greater $\bar{e}$, which is also evident from the shape of the outer contours in the figure. Furthermore, comparing stages (a) and (d), $\bar{e}$ is higher in (a), primarily due to the presence of thermal pressure. Thus, a lower $T_c$, coupled with a higher $\rho_c$, results in reduced stellar deformation owing to increased compactness and decreased thermal pressure. It is worth mentioning that the $\Phi$ also increases with increasing $T_c$.

Visually,  \cref{fig:Tor_isocontous}  shows that although all EoSs produce the signature twin off-axis toroidal lobes of maximum magnetic field strength (red regions), the specific field distribution and overall stellar structure are dependent on the thermodynamic conditions of the EoS. The prolate deformation of the magnetized PNS surface relative to its unmagnetized counterpart is evident in all panels. However, the extent of this deformation, as well as the compactness and location of the high-field toroidal regions, varies significantly across the evolutionary stages represented by the different EoSs. Notably, the cold and catalyzed star (panel (d), $T=0$) appears more compact, with its magnetic field correspondingly confined to a smaller internal volume compared to the hotter, lepton-rich earlier stages at (panels (a)-(b)) and the neutrino transparent stage (panel (c)) this can be attributed to the presence of temperature at these stages that expands the star and reduces its compactness.

All models exhibit a magnetic field distribution characteristic of a purely toroidal configuration: the field strength is null along the symmetry axis, reaches a peak value of $B_{\text{max}} = 5.67\times 10^{17}\,\text{G}$ deep within the stellar core, and gradually decreases to zero at the surface. Despite this common qualitative behavior, notable quantitative differences arise in the structural and global parameters across the various evolutionary phases, as summarized in \cref{tab:table_2}. Among them, the cold and catalyzed NS ($T=0$ model) stands out as the most compact. It possesses the highest central baryon density ($\rho_c = 6.46\times 10^{14}\,\text{g/cm}^3$), the smallest equatorial radius ($r_e = 12.17\,\text{km}$), and the shortest circumferential radius ($R_{\text{circ}} = 14.69\,\text{km}$). This configuration yields a gravitational mass of $1.69\,M_\odot$ and displays a prolate shape, as indicated by a polar-to-equatorial radius ratio of $r_p/r_e = 1.14$ and a deformation parameter of $\bar{e} = -5.00\times 10^{-1}$. Additionally, it exhibits the lowest magnetic flux ($\Phi = 11.24\times 10^{21}\,\text{Wb}$) and the smallest magnetic-to-binding energy ratio ($\mathcal{H}/\mathcal{W} = 1.41\times 10^{-1}$).

In contrast to the cold and catalyzed stellar configuration, the earliest evolutionary stage, characterized by $s_B = 1$ and $Y_l = 0.4$ exhibits a notably less compact structure. It has a lower central baryon density of $\rho_c = 5.88\times 10^{14}\,\text{g/cm}^3$, a larger equatorial radius ($r_e = 16.33\,\text{km}$), and a greater circumferential radius ($R_{\text{circ}} = 18.87\,\text{km}$). Although its gravitational mass is slightly higher at $1.72\,M_\odot$, this configuration also displays a more pronounced prolate deformation, with a radius ratio $r_p/r_e = 1.12$ and a deformation parameter of $\bar{e} = -6.03\times 10^{-1}$. Additionally, it has a higher magnetic flux ($\Phi = 13.59\times 10^{21}\,\text{Wb}$) and a larger $\mathcal{H}/\mathcal{W} = 1.72\times 10^{-1}$ compared to the $T=0$ model. These distinctions highlight the impact of thermal and compositional conditions on the global structure and magnetic properties of PNSs. Even though the first stage is associated with a lower $T_c$ compared to the intermediate stages, it contains a higher neutrino content and is less compact than the cold, compact stellar configuration, as indicated by their $\rho_c$. These properties influence the star's ability to withstand intense magnetic fields. In particular, by comparing stellar deformation to core temperature in \cref{tab:table_2}, we can conclude that temperature plays a significant role: the intermediate stages of stellar evolution, which are associated with higher $T_c$, exhibit greater deformation, as reflected in the higher values of $\bar{e}$.  

Moving to a more evolved but still hot and lepton-rich phase, the configuration with $s_B = 2$ and $Y_l = 0.2$ reveals a further reduction in $\rho_c = 5.35\times 10^{14}\,\text{g/cm}^3$, accompanied by an increase in $r_e = 18.97\,\text{km}$ and $R_{\text{circ}} = 21.49\,\text{km}$ radii. This stage supports a slightly higher gravitational mass of $1.75\,M_\odot$ and displays the most pronounced prolate deformation across all stages, with a radius ratio $r_p/r_e = 1.17$ and deformation parameter $\bar{e} = -9.90\times 10^{-1}$. It also possesses the largest magnetic flux ($\Phi = 17.53\times 10^{21}\,\text{Wb}$) and the highest magnetic-to-binding energy ratio ($\mathcal{H}/\mathcal{W} = 2.30\times 10^{-1}$). The next intermediate state is characterized by $s_B = 2$ after all the neutrinos had escaped from the stellar core, $Y_{\nu_e} = 0$, and it exhibits similar structural properties. At this stage of the star's evolution, the temperature of the stellar matter reaches its peak and begin cool with, $\rho_c = 5.40\times 10^{14}\,\text{g/cm}^3$, equatorial radius of $19.02\,\text{km}$ and circumferential radius of $21.56\,\text{km}$. The gravitational mass slightly decreases to $1.74\,M_\odot$, attributed to the neutrino escape, while the prolate shape remains evident, with $r_p/r_e = 1.16$ and $\bar{e} = -9.36\times 10^{-1}$. The magnetic flux and energy ratio also remain high, with $\Phi = 17.25\times 10^{21}\,\text{Wb}$ and $\mathcal{H}/\mathcal{W} = 2.24\times 10^{-1}$, respectively. These successive comparisons underscore the trend that, as the PNS evolves thermally and deleptonizes, it progressively contracts, exhibits reduced deformation under (see percentage deformation relative to the cold and catalyzed stellar configuration in \cref{tab:pp}) the same magnetic field generation scheme, and confines less magnetic flux, along with a lower magnetic-to-binding energy ratio for the same peak magnetic field strength due to increased compactness and reduced thermal pressure.

In \cref{tab:tp}, we quantify the structural deformations of PNSs at various evolutionary stages when subjected to a purely toroidal magnetic field. As a reference configuration, we adopt the cold, catalyzed NS at $T=0$, studied using the same magnetic field generation formalism. This $T=0$ model, by definition, exhibits zero deformation within our comparative framework and serves as a consistent baseline for evaluating deviations due to thermal and compositional effects. Significant deviations are observed in the first (a) to the third (c) evolutionary stages due to the varied thermodynamic conditions (see references on the study of stellar deformation by magnetic field strength can be found in \cite{Colaiuda:2007br, Mastrano:2015rfa, Mastrano:2013jaa, Franzon:2016iai}). 

For instance, the model with $s_B=2$ and $Y_{\nu_e}=0$ shows equatorial, polar, and volumetric deformations of $56.24\%$, $59.74\%$, and $290.01\%$, respectively, relative to the cold NS reference. A similar degree of deformation is seen in the $s_B=2$, $Y_l=0.2$ model, with increases of $55.72\%$ (equatorial), $60.26\%$ (polar), and $287.30\%$ (volumetric). Even the first stage with $s_B=1$, $Y_l=0.4$, exhibits notable deformations of $34.17\%$, $31.91\%$, and $138.82\%$, respectively. These results highlight the critical role of temperature distribution and compactness in determining the extent of magnetic deformation. As shown in \cref{tab:table_2}, the intermediate stages have substantially higher core temperatures than the initial stage, which correlates with the larger deformations observed. This indicates that the star is more susceptible to magnetic distortion during the deleptonization phase than during the neutrino-trapping stage or in the cold, catalyzed NS configuration. Moreover, the results demonstrate that early-stage PNSs, characterized by elevated entropy and lepton content, are significantly larger and more voluminous than their cold counterparts, despite being modeled under the same magnetic field configuration. This underscores the importance of thermal and compositional effects, in conjunction with the EoS, in shaping the structure and magnetic response of PNSs throughout their evolution \cite{Franzon:2016iai, Lander:2020bou}.

\subsection{PURELY POLOIDAL FIELD}

%\newpage

In this section, we examine the influence of a purely poloidal magnetic field configuration on PNSs across different stages of their evolution. For such magnetic field structures, both the stellar equilibrium and the internal field distribution are governed by the choice of the free function $\mathcal{M}(A_\phi)$, defined in Eq.~(\ref{eq:M_func}). We adopt the linear current approach by setting $\xi = 0$ in this equation, resulting in a poloidal current density given by $J^{\phi} = \rho h k_{\text{pol}}$. Figure~\ref{fig:pol_iso} illustrates the magnetic field strength isocontours ($|B|$) for equilibrium PNS configurations at four key evolutionary stages as discussed thoroughly in previous sections. All models in \cref{fig:pol_iso} consistently exhibit the expected features of a poloidal magnetic configuration: the magnetic field is strongest at the stellar core (red regions), and the field lines (depicted in white) exhibit a predominantly dipolar structure, extending smoothly through the star and into the surrounding space.

\begin{table*}[htbp]
\setlength{\tabcolsep}{3pt}
\renewcommand{\arraystretch}{1.1}

\centering
\begin{tabular}{ccccccccccc}
\hline
Thermodynamic\\ Condition &
\begin{tabular}[c]{@{}c@{}}$\rho_c$\\$(10^{14}\,\mathrm{g/cm^3})$\end{tabular} &
\begin{tabular}[c]{@{}c@{}}$M$\\(M$_\odot$)\end{tabular} &
\begin{tabular}[c]{@{}c@{}}$r_e$\\(km)\end{tabular} &
$r_p/r_e$ &
\begin{tabular}[c]{@{}c@{}}$R_{\text{circ}}$\\(km)\end{tabular} &
\begin{tabular}[c]{@{}c@{}}$\bar{e}$\\$(10^{-1})$\end{tabular} &
\begin{tabular}[c]{@{}c@{}}$\Phi$\\$(10^{21}\,\mathrm{Wb})$\end{tabular} &
\begin{tabular}[c]{@{}c@{}}$\mathcal{H}/\mathcal{W}$\\$(10^{-1})$\end{tabular} &
\begin{tabular}[c]{@{}c@{}}$T_c$\\(MeV)\end{tabular} \\
\hline
$s_B=1,\; Y_l = 0.4$          & 4.93 & 1.69 & 13.80 & 0.76 & 16.50 & 2.40 & 10.38 & 0.96 & 16.49 \\
$s_B=2,\; Y_l = 0.2$          & 4.16 & 1.71 & 14.66 & 0.68 & 17.43 & 2.91 & 11.07 & 1.28 & 35.83 \\
$s_B=2,\; Y_{\nu_e}= 0$       & 4.26 & 1.71 & 14.67 & 0.69 & 17.42 & 2.84 & 11.06 & 1.23 & 37.84 \\
$T=0$                         & 5.57 & 1.67 & 11.44 & 0.78 & 14.13 & 2.04 & 8.66  & 0.77 & 0.0   \\
\hline
\end{tabular}
\caption{Global physical quantities of the equilibrium models with baryonic mass $m_b=1.92\,M_\odot$ and maximum magnetic field strength (purely poloidal) $B_{\max}=5.67\times10^{17}\,\mathrm{G}$. With $\rho_c$ the central baryon density, $M$ the gravitational mass, $r_e$ the equatorial radius, $r_p$ the polar radius, $R_{\text{circ}}$ the circumferential radius, $\bar{e}$ the deformation parameter, $\Phi$ the magnetic flux, $\mathcal{H}$ the total magnetic energy, $\mathcal{W}$ the binding energy, and $T_c$ the core temperature.}
\label{tab:table_pol}
\end{table*}

\begin{figure*}[htbp]
\centering

% Row 1
\includegraphics[width=0.49\textwidth]{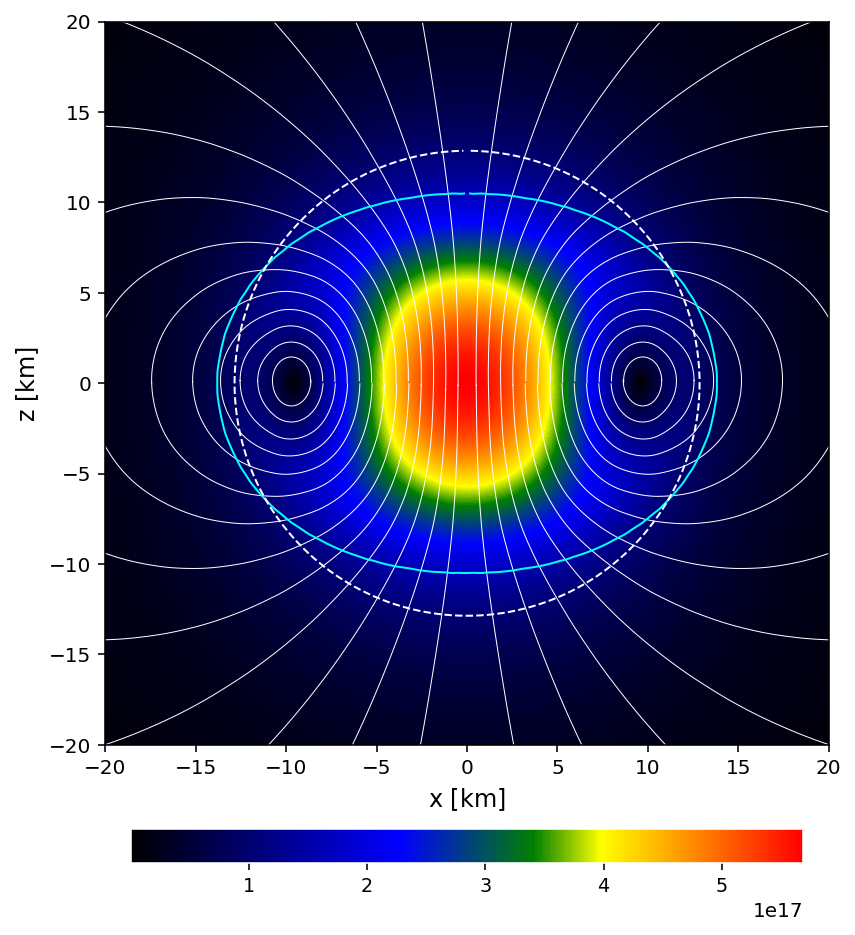}
\hfill
\includegraphics[width=0.49\textwidth]{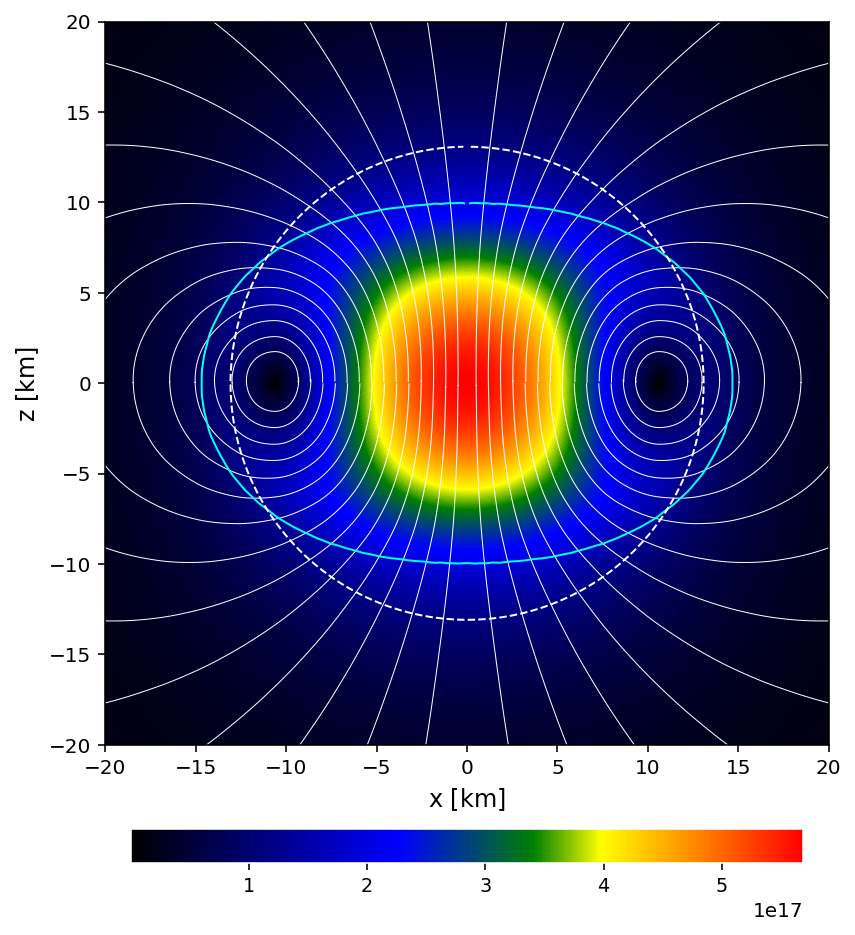}

\vspace{0.5cm}

% Row 2
\includegraphics[width=0.49\textwidth]{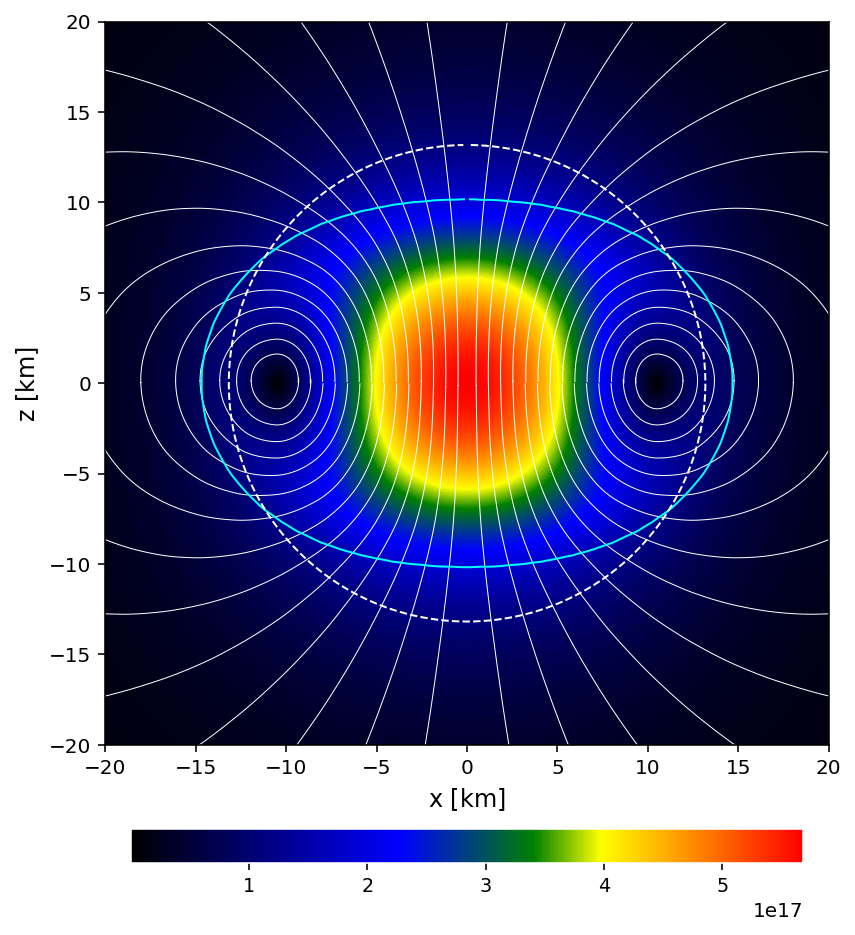}
\hfill
\includegraphics[width=0.49\textwidth]{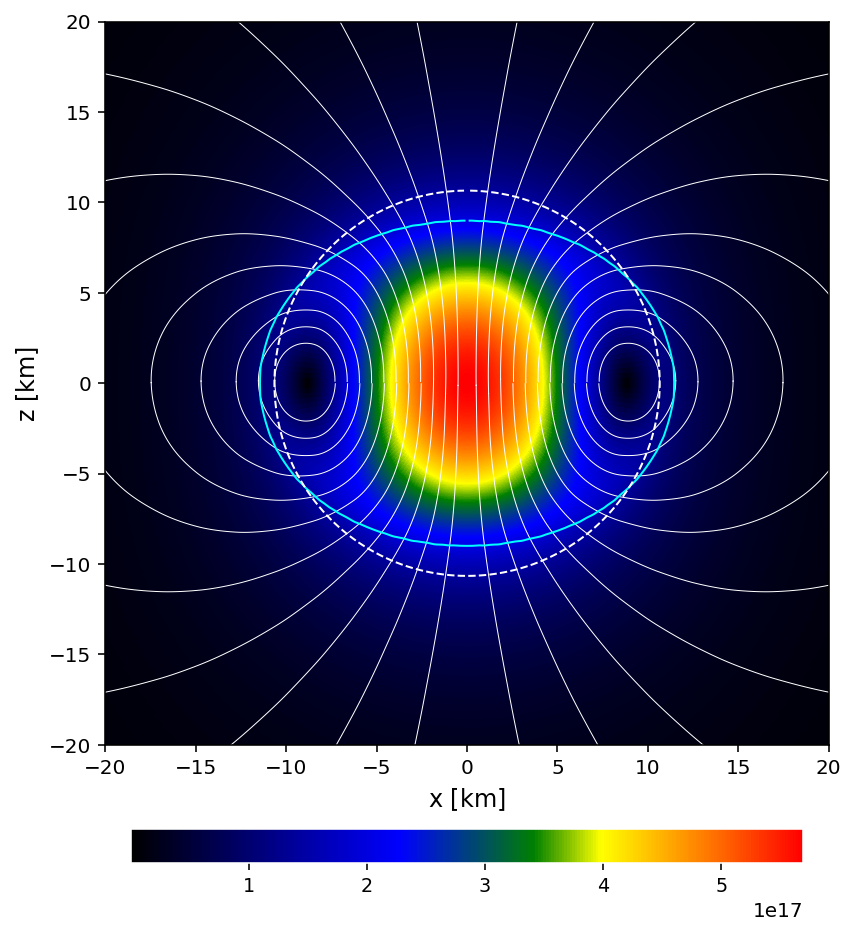}

\caption{Meridional cross-sections of PNSs with a purely poloidal magnetic field at four evolutionary stages: (a) neutrino-trapped ($s_B=1$, $Y_l=0.4$), (b) deleptonizing ($s_B=2$, $Y_l=0.2$), (c) neutrino-transparent ($s_B=2$, $Y_{\nu_e}=0$), and (d) cold and catalyzed ($T=0$). The color map shows the isocontours of magnetic field strength ($|B|$) in Gauss, while the solid cyan and dashed white lines indicate the surfaces of the magnetized and unmagnetized reference stars, respectively. All configurations share the same baryonic mass ($1.92\,M_{\odot}$) and maximum magnetic field ($B_{\text{max}} = 5.67 \times 10^{17}$\,G).}
\label{fig:pol_iso}
\end{figure*}

An oblate deformation of the magnetized stellar surface (solid cyan line) relative to the unmagnetized case (dashed white line) is visible in all panels, indicating compression along the magnetic axis and equatorial extension due to magnetic pressure. The degree of this deformation, overall stellar compactness, and the spatial extent of the high-field regions vary across these thermal and compositional stages. Notably, the cold `mature' NS case in panel (d) is substantially more compact, with the strong magnetic field consequently confined to a smaller volume compared to the hotter, PNS stages, and more radially extended PNS configurations shown in panels (a)--(c). The results presented here are qualitatively in agreement with the ones presented in Refs.~\cite{Pili:2014npa, Franzon:2016iai}

The global physical quantities for equilibrium models at different evolutionary stages of a PNS, maintaining a baryonic mass \(m_b = 1.92 M_\odot\) and a maximum purely poloidal magnetic field strength \(B_{\text{max}} = 5.67 \times 10^{17}\,\text{G}\), are detailed in \cref{tab:table_pol}. Among these, the cold stellar configuration (d), representing a late evolutionary stage stands out as the most compact. It possesses the highest central baryon density, \(\rho_c = 5.57 \times 10^{14}\,\text{g/cm}^3\), the smallest equatorial radius, \(r_e = 11.44\,\text{km}\), and the shortest circumferential radius, \(R_{\text{circ}} = 14.13\,\text{km}\). This configuration yields a gravitational mass of \(1.67\,M_\odot\) and displays an oblate shape, as indicated by a polar-to-equatorial radius ratio of \(r_p/r_e = 0.78\) and a positive deformation parameter of \(\bar{e} = 2.04 \times 10^{-1}\). Additionally, it exhibits the lowest magnetic flux (\( \Phi = 8.66 \times 10^{21}\,\text{Wb}\)) and the smallest magnetic-to-binding energy ratio  (\(\mathcal{H}/\mathcal{W} = 0.77 \times 10^{-1}\)).

In contrast to this late-stage, cold configuration, the first evolutionary stages, characterized by  \(s_B = 1\) and \(Y_l = 0.4\), exhibit a notably less compact structure. It has a lower central baryon density of \(\rho_c = 4.93 \times 10^{14}\,\text{g/cm}^3\), a larger equatorial radius, \(r_e = 13.80\,\text{km}\), and a greater circumferential radius, \(R_{\text{circ}} = 16.50\,\text{km}\). Its gravitational mass is \(1.69\,M_\odot\), and this configuration displays a more pronounced oblate deformation compared to the \(T=0\) mode stage, with a radius ratio \(r_p/r_e = 0.76\) and a deformation parameter of \(\bar{e} = 2.40 \times 10^{-1}\). Furthermore, it has a higher magnetic flux (\( \Phi = 10.38 \times 10^{21}\,\text{Wb}\)) and a larger magnetic-to-binding energy ratio  (\(\mathcal{H}/\mathcal{W} = 0.77 \times 10^{-1}\)), alongside a core temperature of \(16.49\,\text{MeV}\). These distinctions highlight the impact of thermal evolution on the global structure and magnetic properties under strong poloidal fields.

As the PNS evolves further from the first stage described previously, the configuration with \(s_B = 2\) and \(Y_l = 0.2\) represents the deleptoinization phase. This stage, characterized by a higher core temperature of \(35.83\,\text{MeV}\) and increased entropy per baryon, reveals a further reduction in central density, \(\rho_c = 4.16 \times 10^{14}\,\text{g/cm}^3\), accompanied by a significant increase in equatorial, \(r_e = 14.66\,\text{km}\), and circumferential, \(R_{\text{circ}} = 17.43\,\text{km}\), radii. This stage supports a slightly higher gravitational mass of \(1.71\,M_\odot\) and displays the most pronounced oblate deformation across these stages, with a radius ratio \(r_p/r_e = 0.68\) and deformation parameter \(\bar{e} = 2.91 \times 10^{-1}\). It also possesses the largest magnetic flux (\( \Phi = 11.07 \times 10^{21}\,\text{Wb}\)) and the highest magnetic-to-binding energy ratio,  (\(\mathcal{H}/\mathcal{W} = 0.77 \times 10^{-1}\)). These findings underscore the significant role of thermal effects in enhancing both the magnetic flux strength and the resulting stellar deformation during the deleptonization phase.

The third stage, \(s_B = 2, Y_{\nu_e} = 0\), exhibits structural properties that are remarkably similar to the preceding \(s_B=2, Y_l=0.2\) case, yet with subtle distinctions. Its central density, \(\rho_c = 4.26 \times 10^{14}\,\text{g/cm}^3\), is marginally higher, while its equatorial radius (\(14.67\,\text{km}\)) and circumferential radius, \(17.42\,\text{km}\), are almost identical. The gravitational mass remains the same \(1.71\,M_\odot\). The oblate deformation, while still pronounced with \(r_p/r_e = 0.69\) and \(\bar{e} = 2.84 \times 10^{-1}\), is slightly less extreme than in the \(Y_l=0.2\) counterpart. Correspondingly, the magnetic flux (\( \Phi = 11.06 \times 10^{21}\,\text{Wb}\)) and the magnetic-to-binding energy ratio (\(\mathcal{H}/\mathcal{W} = 0.77 \times 10^{-1}\)) are also marginally lower, despite this model achieving the highest core temperature of \(37.84\,\text{MeV}\). This stage marks the peak of thermal evolution before the PNS begins to cool and contract toward its final, cold NS state. Both \(s_B=2\) configurations show significantly more extended and deformed structures compared to the first \(s_B=1\) and the last \(T=0\) stages,  underscoring a clear trend: as the PNS cools and becomes more compact, magnetic deformation and the relative magnetic energy content gradually diminish.

The data in \cref{tab:pp} demonstrate that PNSs in their earlier evolutionary phases exhibit substantial structural deviations compared to the compact cold configuration at $T=0$. For instance, the early evolutionary stage characterized by $s_B=1$ and $Y_l=0.4$ show an equatorial radius, $R_e = 13.81 \text{km}$, a polar radius $R_p = 10.48 \text{km}$, and a volume $V = 8626.74 km^3$. This translates to an equatorial deformation of ${20.70}\%$, a polar deformation of ${16.76}\%$, and a volumetric deformation of ${71.01}\%$ relative to the cold stellar configuration. As the star evolves through the deleptonization and neutrino-transparent phases, represented by $s_B=2, Y_l=0.2$ or $s_B=2, Y_{\nu_e}=0$ respectively, the radii and volume increase further, leading to even more pronounced deformations. %The absolute radii and volume are larger than in the earliest stage, leading to even more pronounced relative deformations. 
Specifically, the $s_B=2, Y_l=0.2$ stage exhibits deformations of ${28.25}\%$ (equatorial), ${10.70}\%$ (polar), and ${87.86}\%$ (volumetric). Similarly, the $s_B=2, Y_{\nu_e}=0$ stage shows comparable equatorial deformation (${28.25}\%$) but a slightly greater polar (${13.19}\%$) and volumetric (${90.74})\%$ increase relative to the cold stellar reference.

\subsection{MIXED FIELD}

Early attempts to model magnetized NSs often employed simplified magnetic field structures—either purely toroidal~\cite{Kiuchi2008, Kiuchi_2009, Frieben:2012dz} or purely poloidal \cite{Bocquet:1995je, Konno_2001, Yaza85.044030}. However, these field configurations were shown to be dynamically unstable~\cite{Tayler/mnras/161.4.365, Wright/mnras/162.4.339, Markey93/mnras/163.1.77}, a fact originally suggested by Prendergast~\cite{pren...123..498P}. More recent simulations have further validated this instability, revealing that magnetic fields with finite helicity naturally evolve toward a stable equilibrium featuring both poloidal and toroidal components~\cite{BraithwaiteId0, Braithwaite2006, Braithwaite.x}. These configurations, known as Twisted Torus (TT) structures, exhibit a toroidal component localized in a ring-shaped region beneath the stellar surface and a poloidal component that extends smoothly into the exterior. Motivated by this, this section focuses on such mixed-field geometries. We model the poloidal field using the same free function $\mathcal{M}(A_\phi)$ as in the purely poloidal case (cf. Eq.~\eqref{eq:M_func}), but restrict it to a linear form by setting $\xi = 0$. The toroidal field is added through the current function $\mathcal{I}(A_\phi)$, where we similarly use a linear profile by setting $\zeta = 0$ and choosing a nonzero amplitude $a \neq 0$. Although both functions are linear, the inclusion of the toroidal component effectively introduces a nonlinear current structure due to its geometry.

\begin{table*}[htbp]
\centering
\scriptsize                         % ← Smaller table entries
\setlength{\tabcolsep}{6pt}
\renewcommand{\arraystretch}{1.15}

\begin{tabular}{cccccccccccc}
\hline
Thermodynamic \\conditions &
$a$ &
\begin{tabular}[c]{@{}c@{}}$\rho_c$\\$(10^{14}\,\mathrm{g/cm^{3}})$\end{tabular} &
\begin{tabular}[c]{@{}c@{}}$M$\\(M$_\odot$)\end{tabular} &
\begin{tabular}[c]{@{}c@{}}$r_e$\\(km)\end{tabular} &
$r_p/r_e$ &
\begin{tabular}[c]{@{}c@{}}$R_{\text{circ}}$\\(km)\end{tabular} &
\begin{tabular}[c]{@{}c@{}}$\bar{e}$\\$(10^{-1})$\end{tabular} &
\begin{tabular}[c]{@{}c@{}}$\Phi$\\$(10^{21}\,\mathrm{Wb})$\end{tabular} &
\begin{tabular}[c]{@{}c@{}}$B_c$\\$(10^{17}\,\mathrm{G})$\end{tabular} &
\begin{tabular}[c]{@{}c@{}}$\mathcal{H}/\mathcal{W}$\\$(10^{-1})$\end{tabular} &
\begin{tabular}[c]{@{}c@{}}$T_c$\\(MeV)\end{tabular} \\
\hline
$s_B=1,\; Y_l = 0.4$         & 0.5 & 4.95 & 1.71 & 14.72 & 0.73 & 17.46 & 2.58 & 11.83 & 4.49 & 1.79 & 16.55 \\
$s_B=2,\; Y_l = 0.2$         & 0.5 & 4.44 & 1.72 & 15.01 & 0.74 & 17.77 & 2.54 & 11.06 & 4.13 & 1.61 & 37.65 \\
$s_B=2,\; Y_{\nu_e} = 0$     & 0.5 & 4.49 & 1.72 & 15.21 & 0.73 & 17.96 & 2.58 & 11.31 & 4.17 & 1.67 & 39.20 \\
$T=0$                        & 0.5 & 5.70 & 1.68 & 11.47 & 0.82 & 14.16 & 1.77 & 8.55  & 4.42 & 0.93 & 0.0   \\
\hline
$s_B=1,\; Y_l = 0.4$         & 1.0 & 5.07 & 1.70 & 15.11 & 0.75 & 17.80 & 2.15 & 10.21 & 4.08 & 1.38 & 16.86 \\
$s_B=2,\; Y_l = 0.2$         & 1.0 & 4.40 & 1.73 & 16.19 & 0.68 & 18.95 & 2.65 & 10.81 & 3.97 & 1.82 & 37.39 \\
$s_B=2,\; Y_{\nu_e} = 0$     & 1.0 & 4.50 & 1.72 & 16.10 & 0.70 & 18.83 & 2.49 & 10.57 & 3.93 & 1.65 & 39.31 \\
$T=0$                        & 1.0 & 5.51 & 1.70 & 12.35 & 0.72 & 15.12 & 2.52 & 10.19 & 4.79 & 1.80 & 0.0   \\
\hline
\end{tabular}
\caption{Global physical quantities of the equilibrium models with baryonic mass $m_b=1.92\,M_\odot$ for mixed-field configurations with $k_{\mathrm{pol}}=0.2$. Here, $\rho_c$ is the central baryon density, $M$ the gravitational mass, $r_e$ the equatorial radius, $r_p$ the polar radius, $R_{\text{circ}}$ the circumferential radius, $\bar{e}$ the deformation parameter, $\Phi$ the magnetic flux, $B_c$ the central magnetic field, $\mathcal{H}$ the total magnetic energy, $\mathcal{W}$ the binding energy, and $T_c$ the core temperature.}
\label{tab:table_mix1}
\end{table*}

We perform a systematic parameter study of mixed-field configurations for a fixed baryonic mass of $1.92\,M_{\odot}$. The analysis is divided into two cases, presented in \cref{tab:table_mix1} and \cref{tab:table_mix2}, to examine the roles of toroidal and poloidal field components separately. In the first case (\cref{tab:table_mix1}), the poloidal magnetization is fixed at $k_{\text{pol}}=0.2$, while the toroidal parameter is varied ($a=0.5,\,1.0$). In the second case (\cref{tab:table_mix2}), the toroidal field is held constant at $a=1.0$, and the poloidal magnetization is varied ($k_{\text{pol}}=0.04,\,0.1$). For each magnetic configuration, we calculate the equilibrium properties at four canonical stages of PNS evolution, providing a detailed understanding of how the PNS structure responds to different magnetic field geometries during cooling and deleptonization.

\begin{figure*}[t]
\centering

% Row 1
\includegraphics[width=0.49\textwidth]{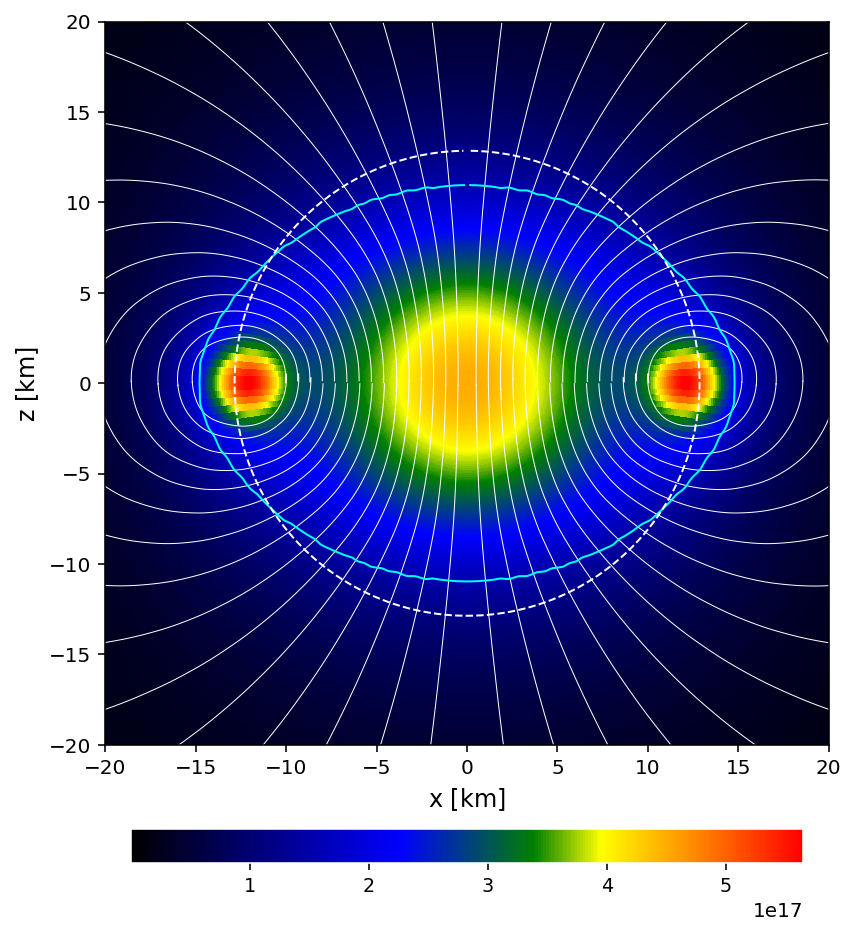}
\hfill
\includegraphics[width=0.49\textwidth]{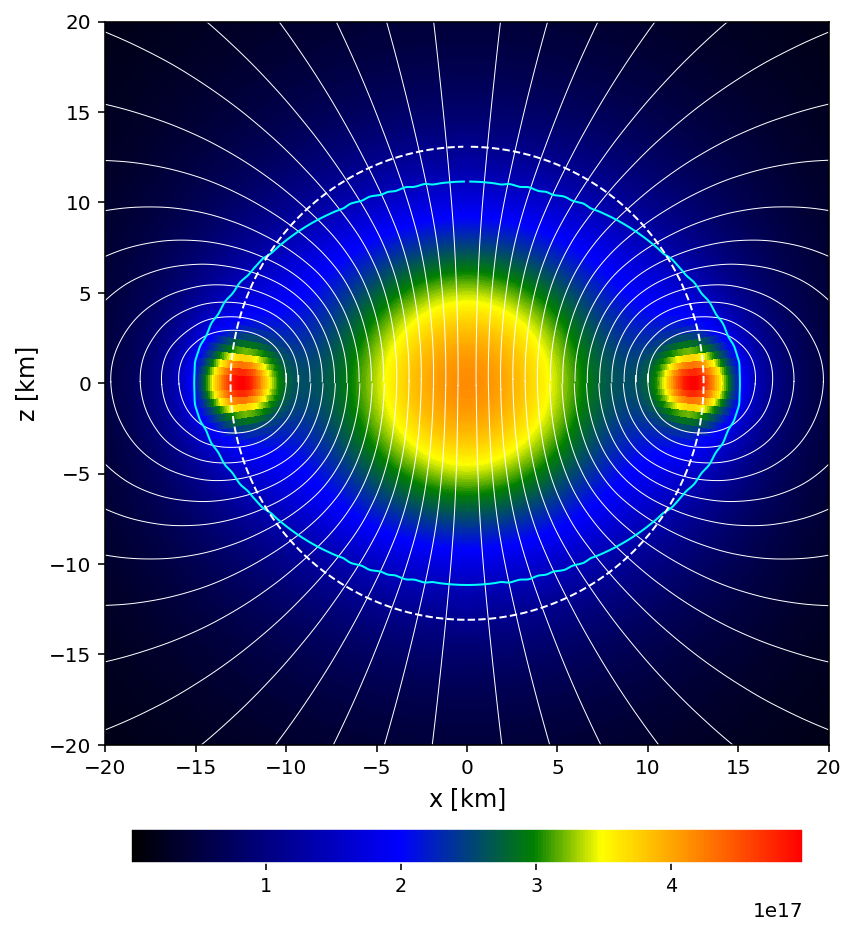}

\vspace{0.5cm}

% Row 2
\includegraphics[width=0.49\textwidth]{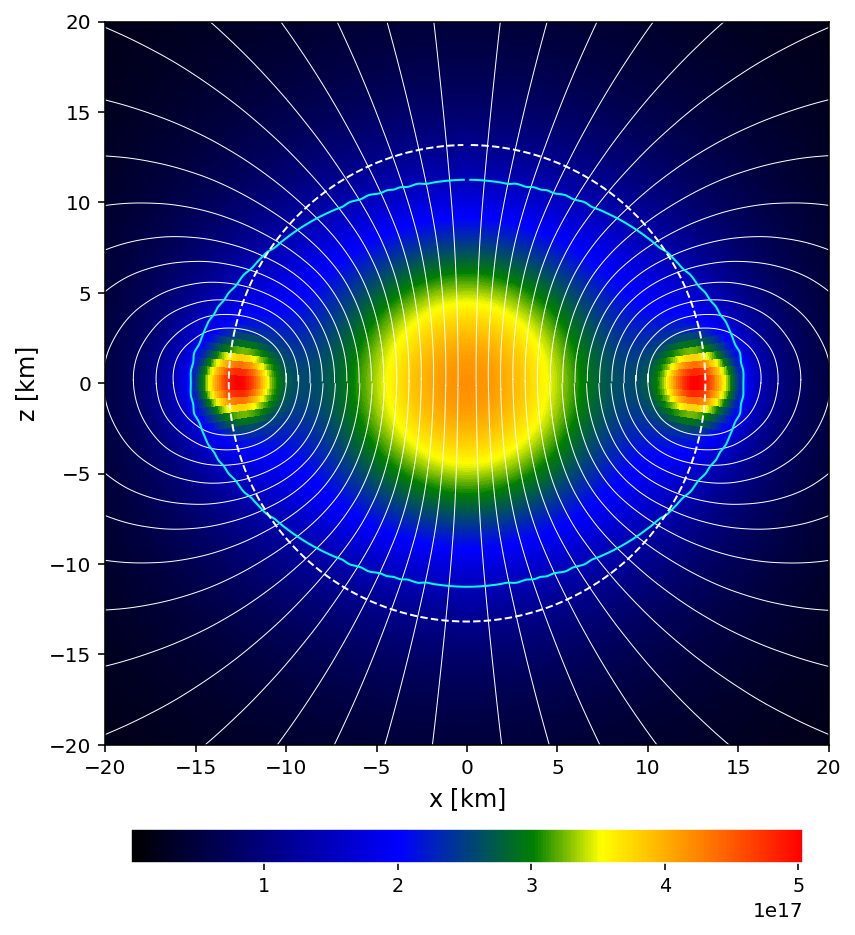}
\hfill
\includegraphics[width=0.49\textwidth]{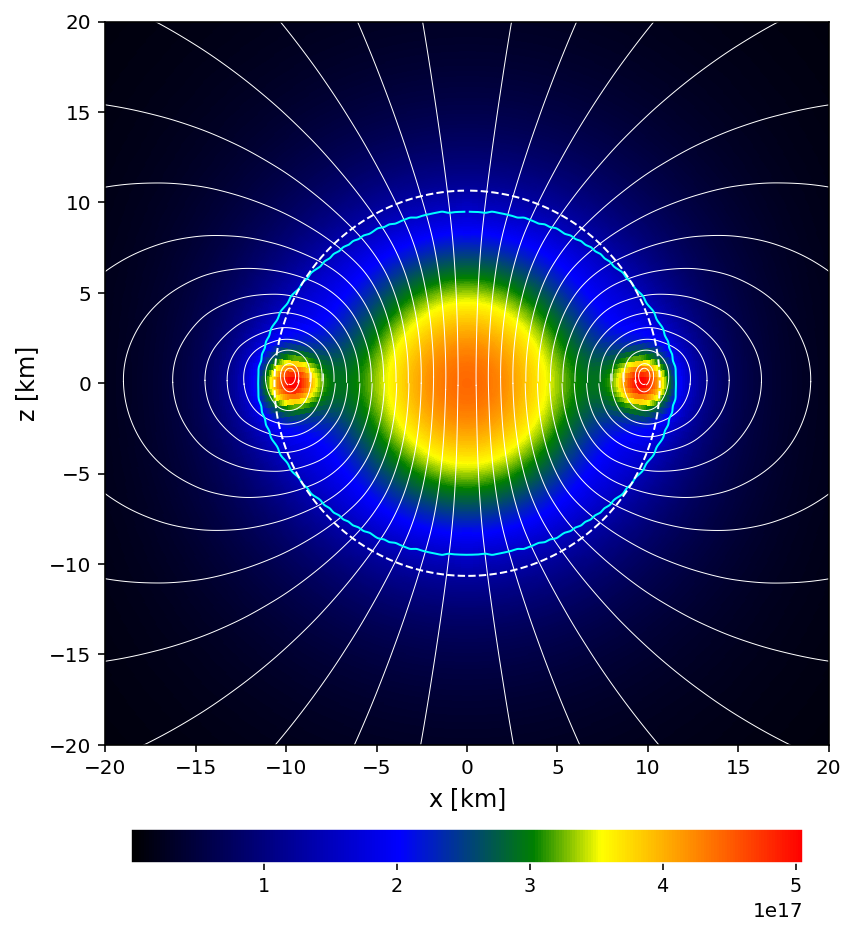}

\caption{Meridional cross-sections of PNSs with a mixed magnetic field at four evolutionary stages: (a)  neutrino-trapped ($s_B=1$, $Y_l=0.4$), (b) deleptonizing ($s_B=2$, $Y_l=0.2$), (c) neutrino-transparent ($s_B=2$, $Y_{\nu_e}=0$), and (d) cold and catalyzed ($T=0$). The color map shows the total magnetic field strength ($|B|$) in Gauss and the white lines trace the poloidal magnetic field lines, while the solid cyan and dashed white lines indicate the surfaces of the magnetized and unmagnetized reference stars, respectively. All configurations share the same baryonic mass ($1.92\,M_{\odot}$) and were computed with magnetic parameters $a=0.5$ and $k_{\text{pol}}=0.2$.}
\label{fig:mixed}
\end{figure*}

Figure~\ref{fig:mixed} presents the magnetic field strength isocontours ($|B|$) for equilibrium PNS configurations at four key evolutionary stages, as previously discussed. These configurations are computed for a fixed baryonic mass and magnetization parameters ($a = 0.5$, $k_{\text{pol}} = 0.2$). The poloidal magnetic field structure closely follows that seen in purely poloidal models: it permeates the entire PNS, reaches its maximum at the center, and vanishes in a ring-like region along the equatorial plane. Outside the PNS surface, the field transitions smoothly and exhibits a predominantly dipolar structure. In contrast, the toroidal magnetic field shows a distinctly different behavior. It does not occupy the full interior but is instead confined to a toroidal region near the equator, where it attains its peak strength -- coinciding with the location where the poloidal component vanishes. This configuration results from the specific choice of the poloidal current distribution and its confinement within the stellar volume. Across all panels, the solid cyan line outlines the oblate deformation of the magnetized PNS surface relative to the unmagnetized case (dashed white line). The thermodynamic and compositional evolution of the PNS directly influences its global structure, modifying its compactness, surface shape, and the spatial extent of strong magnetic fields. In panel~(d), the final cold and mature NS is significantly more compact than the hotter PNS stages shown in panels~(a) -- (c), leading to a more localized confinement of the high-field region.

To examine the structural effects of the toroidal magnetic field component, we compare the equilibrium configurations listed in \cref{tab:table_mix1} for two cases: a moderate toroidal field ($a = 0.5$) and a strong toroidal field ($a = 1.0$), while keeping the poloidal magnetization fixed at $k_{\text{pol}} = 0.2$. The impact of increasing toroidal field strength on the stellar structure does not follow a simple monotonic trend but instead exhibits a strong dependence on the PNS evolutionary stage, as discussed in detail below.

\begin{table*}[t]
\centering
\scriptsize                         % Smaller but readable
\setlength{\tabcolsep}{6pt}        % Column spacing
\renewcommand{\arraystretch}{1.15} % Vertical spacing

\begin{tabular}{cccccccccccc}
\hline
Thermodynamic\\ conditions &
$k_{\mathrm{pol}}$ &
\begin{tabular}[c]{@{}c@{}}$\rho_c$\\$(10^{14}\,\mathrm{g/cm^{3}})$\end{tabular} &
\begin{tabular}[c]{@{}c@{}}$M$\\(M$_\odot$)\end{tabular} &
\begin{tabular}[c]{@{}c@{}}$r_e$\\(km)\end{tabular} &
$r_p/r_e$ &
\begin{tabular}[c]{@{}c@{}}$R_{\text{circ}}$\\(km)\end{tabular} &
\begin{tabular}[c]{@{}c@{}}$\bar{e}$\\$(10^{-1})$\end{tabular} &
\begin{tabular}[c]{@{}c@{}}$\Phi$\\$(10^{21}\,\mathrm{Wb})$\end{tabular} &
\begin{tabular}[c]{@{}c@{}}$B_c$\\$(10^{17}\,\mathrm{G})$\end{tabular} &
\begin{tabular}[c]{@{}c@{}}$\mathcal{H}/\mathcal{W}$\\$(10^{-1})$\end{tabular} &
\begin{tabular}[c]{@{}c@{}}$T_c$\\(MeV)\end{tabular} \\
\hline
$s_B=1,\; Y_l = 0.4$         & 0.04 & 5.40 & 1.67 & 12.84 & 0.99 & 15.43 & 0.07 & 1.78 & 0.78 & 0.03 & 17.70 \\
$s_B=2,\; Y_l = 0.2$         & 0.04 & 4.88 & 1.68 & 13.14 & 0.98 & 15.74 & 0.08 & 1.80 & 0.75 & 0.04 & 40.46 \\
$s_B=2,\; Y_{\nu_e} = 0$     & 0.04 & 4.95 & 1.68 & 13.24 & 0.98 & 15.83 & 0.08 & 1.79 & 0.74 & 0.03 & 41.84 \\
$T=0$                        & 0.04 & 5.91 & 1.65 & 10.68 & 0.99 & 13.27 & 0.09 & 1.85 & 0.92 & 0.05 & 0.0   \\
\hline
$s_B=1,\; Y_l = 0.4$         & 0.10 & 5.46 & 1.67 & 13.24 & 0.94 & 15.84 & 0.51 & 4.70 & 2.00 & 0.25 & 17.84 \\
$s_B=2,\; Y_l = 0.2$         & 0.10 & 4.95 & 1.68 & 13.54 & 0.93 & 16.16 & 0.59 & 4.85 & 1.94 & 0.28 & 40.87 \\
$s_B=2,\; Y_{\nu_e} = 0$     & 0.10 & 5.01 & 1.68 & 13.63 & 0.94 & 16.26 & 0.57 & 4.83 & 1.93 & 0.27 & 42.20 \\
$T=0$                        & 0.10 & 5.97 & 1.66 & 10.98 & 0.93 & 13.60 & 0.60 & 4.65 & 2.29 & 0.30 & 0.0   \\
\hline
\end{tabular}
\caption{Global physical quantities of the equilibrium models with baryonic mass $m_b=1.92\,M_\odot$ for mixed-field configurations with $a=1.0$. Here, $\rho_c$ is the central baryon density, $M$ the gravitational mass, $r_e$ the equatorial radius, $r_p$ the polar radius, $R_{\text{circ}}$ the circumferential radius, $\bar{e}$ the deformation parameter, $\Phi$ the magnetic flux, $B_c$ the central magnetic field, $\mathcal{H}$ the total magnetic energy, $\mathcal{W}$ the binding energy, and $T_c$ the core temperature.}
\label{tab:table_mix2}
\end{table*}

At the early,  ($s_B = 1$, $Y_l = 0.4$), increasing the toroidal magnetic field strength from $a = 0.5$ to $a = 1.0$ leads to a reduction in stellar oblateness. This is evident from the decrease in the deformation parameter, $\bar{e}$, from $2.58 \times 10^{-1}$ to $2.15 \times 10^{-1}$, along with a slight increase in the polar-to-equatorial radius ratio ($r_p/r_e$) from 0.73 to 0.75. The star also becomes more radially extended, with the equatorial radius $r_e$ increasing from $14.72\,\mathrm{km}$ to $15.11\,\mathrm{km}$, and the circumferential radius $R_{\text{circ}}$ growing from $17.46\,\mathrm{km}$ to $17.80\,\mathrm{km}$. This expansion is accompanied by a decrease in the central magnetic field strength, $B_c$, from $4.49 \times 10^{17}\,\mathrm{G}$ to $4.08 \times 10^{17}\,\mathrm{G}$, and a reduction in the total magnetic flux, $ \Phi$, from $11.83 \times 10^{21}\,\mathrm{Wb}$ to $10.21 \times 10^{21}\,\mathrm{Wb}$. Consequently, the magnetic-to-binding energy ratio, $\mathcal{H/W}$, drops from $1.79 \times 10^{-1}$ to $1.38 \times 10^{-1}$. The gravitational mass, $M$, decreases slightly from $1.71\,M_{\odot}$ to $1.70\,M_{\odot}$, while the central density, $\rho_c$, and core temperature, $T_c$, increase marginally.

In a significant departure from the star's birth, $s_B=2, Y_l=0.2$, strengthening the toroidal field during this deleptonizing phase makes the star more oblate. The deformation parameter $\bar{e}$ increases from $2.54 \times 10^{-1}$ to $2.65 \times 10^{-1}$, and the radius ratio $r_p/r_e$ drops sharply from 0.74 to 0.68. This is accompanied by a substantial increase in both the equatorial radius from $15.01\,$km to $16.19\,$km and the circumferential radius from $17.77\,$km to $18.95\,$km. The central density decreases from $4.44 \times 10^{14}\,\mathrm{g/cm}^3$ to $4.40 \times 10^{14}\,\mathrm{g/cm}^3$, while the gravitational mass shows a slight increase from $1.72\,M_{\odot}$ to $1.73\,M_{\odot}$. The core temperature is slightly lower for the stronger field, at $37.39\,$MeV compared to $37.65\,$MeV. While the central field strength and magnetic flux decrease slightly, the overall magnetic energy relative to the binding energy $\mathcal{H/W}$ increases from $1.61 \times 10^{-1}$ to $1.82 \times 10^{-1}$, indicating a complex redistribution of energy during deleptonization.

At the neutrino-transparent stage ($s_B = 2$, $Y_{\nu_e} = 0$), the effects of the toroidal magnetic field resemble those observed at the initial, lepton-rich stage. A stronger toroidal field ($a = 1.0$) again leads to reduced stellar oblateness, with the deformation parameter $\bar{e}$ decreasing from $2.58 \times 10^{-1}$ to $2.49 \times 10^{-1}$. The star remains more radially extended in the presence of the stronger field, with the equatorial radius increasing from $15.21\,\mathrm{km}$ ($a = 0.5$) to $16.10\,\mathrm{km}$ ($a = 1.0$). In contrast, the gravitational mass and central density show negligible change between the two cases. The core temperature is slightly higher for the stronger field, increasing from $39.20\,\mathrm{MeV}$ to $39.31\,\mathrm{MeV}$. Consistent with other hot evolutionary stages, the central magnetic field strength decreases as the toroidal field strengthens, dropping from $4.17 \times 10^{17}\,\mathrm{G}$ to $3.93 \times 10^{17}\,\mathrm{G}$.

The final, cold, Catalyzed NS ($T=0$) state exhibits the most pronounced and often reversed response to the toroidal field. Here, a stronger toroidal component makes the star more oblate, with the deformation parameter $\bar{e}$ increasing from $1.77 \times 10^{-1}$ to $2.52 \times 10^{-1}$. The star is both more massive ($1.70\,M_{\odot}$ vs.\ $1.68\,M_{\odot}$) and has a significantly larger equatorial radius ($12.35\,$km vs.\ $11.47\,$km). In a key reversal of the trend observed in the hot PNS stages, both the central magnetic field strength and the overall magnetic energy are substantially higher for the stronger toroidal field. $B_c$ increases from $4.42 \times 10^{17}\,\mathrm{G}$ to $4.79 \times 10^{17}\,\mathrm{G}$, and the $\mathcal{H/W}$ ratio nearly doubles from $0.93 \times 10^{-1}$ to $1.80 \times 10^{-1}$. This highlights that the toroidal field's structural role is fundamentally different and more pronounced in a compact, gravitationally dominated object than in a thermally-supported PNS.

We now turn to the structural impact of the poloidal magnetic field component, with the results presented in \cref{tab:table_mix2}. In this analysis, we hold the strong toroidal field constant ($a=1.0$) and compare models with a weak poloidal magnetization ($k_{pol}=0.04$) to those with a stronger one ($k_{pol}=0.1$). Unlike the complex behavior seen when varying the toroidal field, the response to a stronger poloidal field is consistent across all evolutionary stages.

From the initial stage, ($s_B=1, Y_l=0.4$), increasing the poloidal field strength from $k_{pol}=0.04$ to $k_{pol}=0.1$ makes the star more oblate. The deformation parameter, $\bar{e}$, increases by a factor of seven, from a nearly spherical $0.07 \times 10^{-1}$ to a highly deformed $0.51 \times 10^{-1}$. This is mirrored by the radius ratio $r_p/r_e$ dropping from 0.99 to 0.94. The star also becomes slightly larger, with the equatorial radius $r_e$ increasing from $12.84\,$km to $13.24\,$km and the circumferential radius $R_{\text{circ}}$ growing from $15.43\,$km to $15.84\,$km. The central density $\rho_c$ rises from $5.40 \times 10^{14}\,\mathrm{g/cm}^3$ to $5.46 \times 10^{14}\,\mathrm{g/cm}^3$, while the gravitational mass $M$ remains unchanged at $1.67\,M_{\odot}$. The core temperature $T_c$ also increases slightly from $17.70\,$MeV to $17.84\,$MeV. As expected, all magnetic quantities increase substantially: the total magnetic flux $ \Phi$ rises from $1.78 \times 10^{21}\,$Wb to $4.70 \times 10^{21}\,$Wb, the central field $B_c$ increases from $0.78 \times 10^{17}\,\mathrm{G}$ to $2.00 \times 10^{17}\,\mathrm{G}$, and the magnetic energy ratio ($\mathcal{H/W}$) grows from $0.03 \times 10^{-1}$ to $0.25 \times 10^{-1}$.

This consistent trend continues during the hot deleptonization phase $s_B=2, Y_l=0.2$. A stronger poloidal field again produces a significantly more oblate star, with $\bar{e}$ increasing from $0.08 \times 10^{-1}$ to $0.59 \times 10^{-1}$ and the radius ratio ($r_p/r_e$) decreasing from 0.98 to 0.93. The equatorial radius expands from $13.14\,$km to $13.54\,$km, and the circumferential radius increases from $15.74\,$km to $16.16\,$km. The central density rises from $4.88 \times 10^{14}\,\mathrm{g/cm}^3$ to $4.95 \times 10^{14}\,\mathrm{g/cm}^3$, with the gravitational mass remaining constant at $1.68\,M_{\odot}$. The core temperature is also higher for the stronger field, rising from $40.46\,$MeV to $40.87\,$MeV. The magnetic properties are again enhanced across the board: $ \Phi$ increases from $1.80 \times 10^{21}\,$Wb to $4.85 \times 10^{21}\,$Wb, $B_c$ rises from $0.75 \times 10^{17}\,\mathrm{G}$ to $1.94 \times 10^{17}\,\mathrm{G}$, and the $\mathcal{H/W}$ ratio increases sevenfold from $0.04 \times 10^{-1}$ to $0.28 \times 10^{-1}$.

The behavior persists in the neutrino-transparent phase ($s_B=2, Y_{\nu_e}=0$). Strengthening the poloidal component results in a much more oblate star, with $\bar{e}$ jumping from $0.08 \times 10^{-1}$ to $0.57 \times 10^{-1}$ and the radius ratio $r_p/r_e$ dropping from 0.98 to 0.94. The star is slightly larger, with $r_e$ increasing from $13.24\,$km to $13.63\,$km and $R_{\text{circ}}$ from $15.83\,$km to $16.26\,$km. The central density increases from $4.95 \times 10^{14}\,\mathrm{g/cm}^3$ to $5.01 \times 10^{14}\,\mathrm{g/cm}^3$ while the mass remains unchanged at $1.68\,M_{\odot}$. The core temperature also increases from $41.84\,$MeV to $42.20\,$MeV. All magnetic quantities -- $ \Phi$, $B_c$, and $\mathcal{H/W}$ -- show a large and consistent increase, with values for $k_{pol}=0.1$ being several times larger than those of $k_{pol}$, reaffirming the dominant role of the poloidal component.

The final cold, Catalyzed NS state exhibits the same qualitative response. A stronger poloidal field leads to a highly oblate configuration, with $\bar{e}$ increasing from $0.09 \times 10^{-1}$ to $0.60 \times 10^{-1}$ and the radius ratio $r_p/r_e$ decreasing from 0.99 to 0.93. The equatorial radius expands from $10.68\,$km to $10.98\,$km, and the circumferential radius grows from $13.27\,$km to $13.60\,$km. The star is slightly more massive $1.66\,M_{\odot}$ vs.\ $1.65\,M_{\odot}$ and has a higher central density $5.97 \times 10^{14}\,\mathrm{g/cm}^3$ vs.\ $5.91 \times 10^{14}\,\mathrm{g/cm}^3$. The magnetic properties see the greatest enhancement in this compact state: $ \Phi$ increases from $1.85 \times 10^{21}\,$Wb to $4.65 \times 10^{21}\,$Wb, $B_c$ more than doubles from $0.92 \times 10^{17}\,\mathrm{G}$ to $2.29 \times 10^{17}\,\mathrm{G}$, and the $\mathcal{H/W}$ ratio increases from $0.05 \times 10^{-1}$ to $0.30 \times 10^{-1}$. This behavior is physically expected. As the star cools and contracts, as indicated by the decreasing stellar radius relative to hotter stages, magnetic flux conservation $\bigl( \Phi = B_c \times \text{area}\bigr)$ requires the central magnetic field $B_c$ to increase to maintain approximately constant magnetic flux \cite{Lai:2000at, Lattimer:2006xb}, as shown in \cref{tab:table_mix2}. Additionally, in the first stage, when the star is relatively compact due to lower temperature and higher $Y_l$, we observe higher values of $B_c$ compared to the less compact intermediate stages, when the star remains relatively hotter.  This uniform response confirms that the poloidal field is the primary driver of oblateness and magnetic energy in these mixed-field models.

\subsection{Magnetic decay time}

\begin{figure*}[htbp]
  \centering
  \begin{minipage}[b]{0.78\linewidth}
    \centering
    \includegraphics[width=\linewidth]{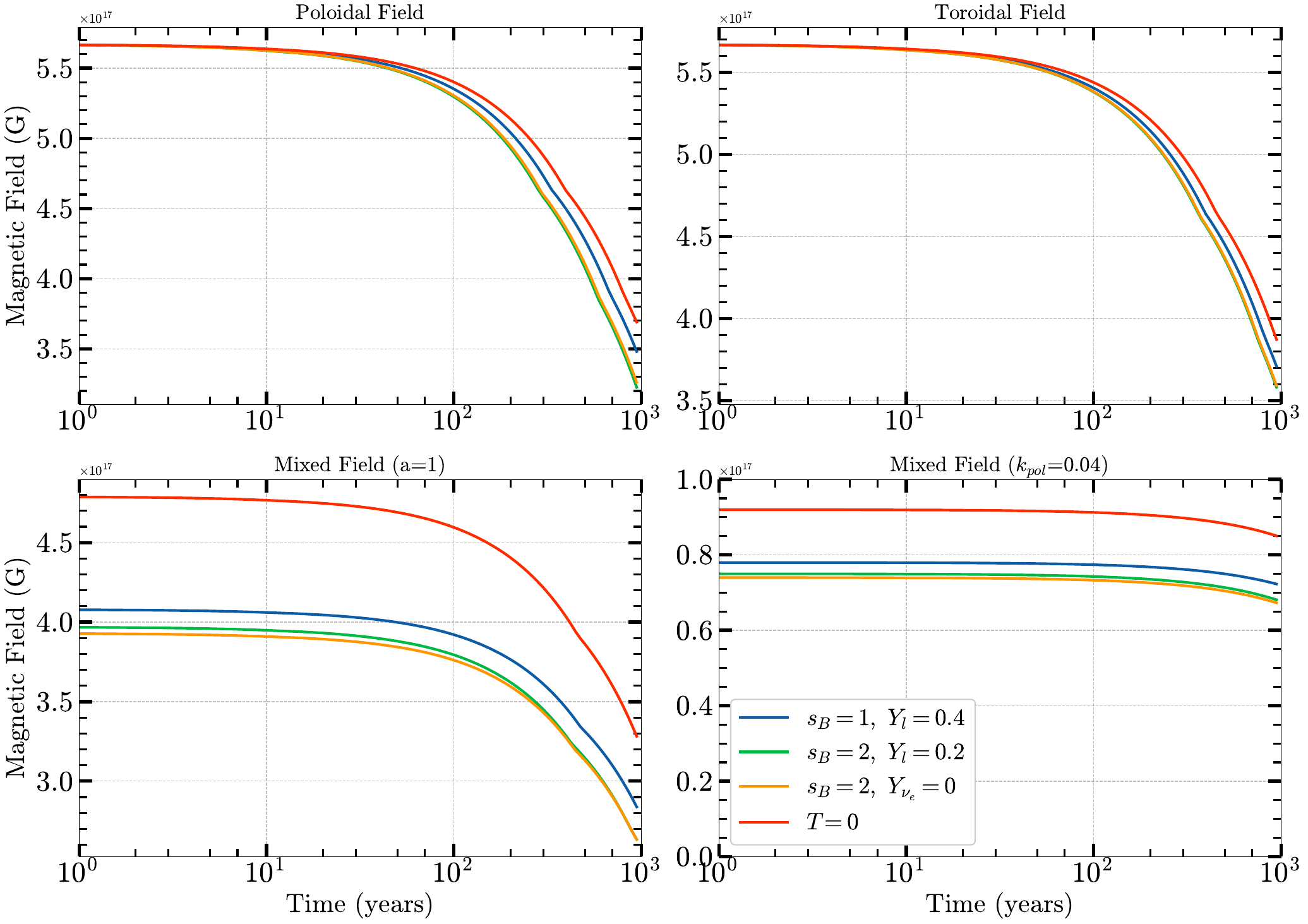 }
  \end{minipage}\hspace{0.02\linewidth}%
  \caption{Time evolution of the maximum magnetic field strength for different magnetic field configurations in PNSs. The upper left and upper right panels show the decay of the poloidal and toroidal magnetic fields, respectively, while the lower panels depict the evolution of mixed field configurations: the lower left for $a = 1$ and the lower right for $k_{pol} = 0.04$. Each curve represents a distinct  PNS model: $s_B=1$, $Y_l=0.4$ (blue), $s_B=2$, $Y_l=0.2$ (green), $s_B=2$, $Y_{\nu_e}=0$ (orange), and a cold deleptonized $T=0$ model (red).}
  \label{fig:magnetic_decay}
\end{figure*}

The long-term evolution of magnetic fields in PNSs and mature NSs is governed by three primary dissipative processes: Ohmic dissipation, ambipolar diffusion, and Hall drift. The dominant mechanism at any time depends on local conditions such as magnetic field strength, central density, and temperature, with each process characterized by a distinct timescale~\cite{goldreich}.

The characteristic decay timescale for Ohmic dissipation, $\tau_{\text{Ohmic}}$, is given by:

\begin{equation}
\tau_{\text{Ohmic}} \sim 2 \times 10^{11} \left(\frac{L_5^2}{T_8^2}\right) \left(\frac{\rho}{\rho_{\text{nuc}}}\right)^{3}  \text{yr}.
\end{equation}
The timescale for ambipolar diffusion, \(t_{\text{ambipolar}}\), can be expressed as:

\begin{equation*}
\tau^s_{\text{ambipolar}} \sim 3 \times 10^9 \left(\frac{L_5^2\times T_8^2}{B_{12}^2}\right) \;  \text{yr} ,\\
\end{equation*}
\begin{equation}
   \tau_{ambipolar} \sim \frac{5\times10^{15}}{T_8^6\times B_{12}^2}\;yr\;+\tau^s_{\text{ambipolar}},
\end{equation} 
The Hall drift timescale, \(t_{\text{Hall}}\), is:

\begin{equation}
\tau_{\text{Hall}} \sim 5 \times 10^8 \left(\frac{L_5^2}{B_{12}}\right) \left(\frac{\rho}{\rho_{\text{nuc}}}\right)\;  \text{yr}.
\end{equation}
Here, $L_5$ is the characteristic length scale of flux loops in units of $10^5$~cm; $T_8$ is the core temperature in units of $10^8$~K; and $B_{12}$ is the field strength in units of $10^{12}$~G. It is important to note that for the `cold' ($T=0$) model, we use a representative low temperature of 0.1~MeV. This is equivalent to approximately $1.16 \times 10^9$~K, corresponding to a value of $T_8 \approx 11.6$ in our calculations. This choice is considered reasonable, as the temperature in cold and catalyzed NS is typically estimated to be $T< 1$ MeV \cite{Prakash:1996xs}. The dominant magnetic decay mechanism transitions with field strength: from Ohmic decay for weak fields ($B \lesssim 10^{11}$ G), to Hall drift for intermediate fields ($B \sim 10^{12}-10^{13}$ G), and finally to rapid ambipolar diffusion in the strong-field regime ($B \gtrsim 10^{14}$ G)\cite{Heyl,goldreich}.

The decay of magnetic fields in NS can be studied by solving the decay equation as described in \cite{Heyl},
\begin{equation}
    \frac{dB}{dt} = -B(t) \left( \frac{1}{\tau_{ohmic}} + \frac{1}{\tau_{ambipolar}} + \frac{1}{\tau_{hall}} \right)
\end{equation}

Figure~\ref{fig:magnetic_decay} shows the magnetic field evolution over $1000\,\mathrm{yr}$ for purely poloidal, purely toroidal, and mixed-field configurations across the four thermodynamic stages of PNS evolution. For the pure-field cases, all models are initialized with the same peak field strength, $B_{\max}=5.67\times10^{17}\,\mathrm{G}$, enabling a direct comparison of dissipation efficiency. In both poloidal and toroidal geometries, magnetic decay is strongly correlated with the thermal state of the star, with the largest fractional losses occurring during the hot, high-entropy deleptonizing and neutrino-transparent stages. In these phases, the field strength is reduced by $\sim40\%$ for poloidal fields and $\sim37\%$ for toroidal fields, whereas the cold, catalyzed configuration exhibits the most suppressed decay, with losses below $\sim35\%$. The neutrino-trapped stage displays intermediate behavior.

Mixed-field configurations exhibit the same qualitative trend despite differences in their initial field strengths, which depend on the underlying thermodynamic state. For the fiducial mixed model ($a=1$, $k_{\mathrm{pol}}=0.2$), the hottest stages show the most efficient decay, with field reductions of $\sim33\%$, while the cold and neutrino-trapped configurations experience more moderate losses of $\sim30$--$32\%$. A weaker mixed-field configuration ($k_{\mathrm{pol}}=0.04$) shows smaller overall decay, but the thermal ordering persists, with high-entropy models decaying more efficiently ($\sim9\%$) than colder ones ($\lesssim8\%$).

These calculations assume a fixed core temperature for each evolutionary stage and are not intended to track the detailed cooling of an individual PNS. Instead, they isolate the intrinsic efficiency of magnetic dissipation mechanisms across different thermodynamic states. The consistent enhancement of decay in hot, high-entropy configurations across all magnetic geometries demonstrates that temperature-driven processes dominate the early evolution of large-scale magnetic fields, reinforcing the conclusion that the thermodynamic state at birth plays a decisive role in shaping the early magnetic evolution of neutron stars.

\section{Final Remarks and Conclusion}\label{conclusion}

In this work, we present a systematic general-relativistic study of the structural evolution of strongly magnetized protoneutron stars. Using the \texttt{XNS 4.0} code and isentropic EoS derived from density-dependent relativistic mean-field theory (DDME2), we construct self-consistent equilibrium models in charge-neutral, $\beta$-equilibrated matter for purely toroidal, purely poloidal, and mixed magnetic field configurations. These models are followed across four representative thermodynamic stages of PNS evolution: neutrino-trapped, deleptonizing, neutrino-transparent, and the final cold, catalyzed NSs.

Our results demonstrate that the magnetic response of a PNS is primarily controlled by its thermal and compositional state. Magnetic effects are strongest during the intermediate, high-entropy deleptonizing and neutrino-transparent phases, where reduced compactness renders the star more deformable and allows stronger flux confinement. In contrast, the cold, catalyzed configuration is maximally compact and correspondingly more rigid, exhibiting minimal deformation and a reduced magnetic energy content.

By constructing sequences at fixed peak magnetic field strength, $B_{\max}$, we isolate the structural response of the star to changing thermodynamic conditions without assuming magnetic flux conservation. This approach is well justified during the Kelvin--Helmholtz phase ($\sim$10--20 s), when thermal and lepton evolution proceeds slowly compared to hydrodynamic and Alfv\'en timescales. We find that magnetic geometry plays a decisive role: toroidal fields induce prolate deformations, poloidal fields produce oblate configurations, and mixed (twisted-torus) geometries yield stable intermediate behavior. For a fixed baryonic mass of $1.92\,M_\odot$, hot PNSs can exhibit volumetric deformations up to nearly three times larger than their cold counterparts under the same toroidal field strength.

We further estimate magnetic field decay timescales due to Ohmic dissipation, Hall drift, and ambipolar diffusion, finding a strong sensitivity to core temperature. Hotter PNS phases undergo significantly faster magnetic dissipation, indicating that temperature-enhanced decay dominates the early magnetic evolution of NSs largely independent of field geometry.

Overall, this study highlights the coupled roles of thermodynamics, magnetic topology, and stellar structure in shaping the evolution of magnetized NSs from birth to maturity. Realistic modeling of PNSs must therefore account for both thermal evolution and magnetic field geometry, as they critically influence observable properties such as stellar deformation, magnetic energy content, and the long-term survival of strong magnetic fields. While our use of fixed magnetic parameters provides a controlled baseline, future work incorporating magnetic flux conservation is expected to reveal additional field amplification during cooling and contraction, further enriching the magnetic evolution of young neutron stars.

\acknowledgments

A.I. acknowledges financial support from the São Paulo State Research Foundation (FAPESP), Grant No. 2023/09545-1. This work is part of the project INCT-FNA (Proc. No. 464898/2014-5). P.~Thakur is supported by the National Research Foundation of Korea (NRF) grant funded by the Korea government (MSIT) (No.~RS-2024-00457037). This work was supported (in part) by the Yonsei University Research Fund(Yonsei University Frontier Fellowship for Postdoctoral Researchers) of 2025.

\newpage

\appendix
\section{The equatorial, polar, and volumetric deformation for all three magnetic field configurations}

\begin{table*}[htbp]
\centering
\scriptsize
\setlength{\tabcolsep}{6pt}
\renewcommand{\arraystretch}{1.3}

\caption{Equatorial, polar and volumetric deformations of each EOS with respect to the $T=0$ configuration for a purely toroidal magnetic field.}
\label{tab:tp}

\begin{tabular}{ccccccc}
\hline
Thermodynamic \\ conditions &
\begin{tabular}{c}
$R_{e}$ \\
(km)
\end{tabular} &
\begin{tabular}{c}
$R_{p}$ \\
(km)
\end{tabular} &
\begin{tabular}{c}
Volume \\
(km$^3$)
\end{tabular} &
\begin{tabular}{c}
Equatorial \\
deformation (\%)
\end{tabular} &
\begin{tabular}{c}
Polar \\
deformation (\%)
\end{tabular} &
\begin{tabular}{c}
Volumetric \\
deformation (\%)
\end{tabular} \\
\hline
$s_B=1,\;Y_l=0.4$     & 16.33 & 18.25 & 20394.31 & 34.17 & 31.91 & 138.82 \\
$s_B=2,\;Y_l=0.2$     & 18.96 & 22.18 & 33074.27 & 55.72 & 60.26 & 287.30 \\
$s_B=2,\;Y_{\nu_e}=0$ & 19.02 & 22.10 & 33305.50 & 56.24 & 59.74 & 290.01 \\
$T=0$                 & 12.17 & 13.84 &  8539.66 &  0.00 &  0.00 &   0.00 \\
\hline
\end{tabular}
\end{table*}

\begin{table*}[htbp]
\centering
\scriptsize
\setlength{\tabcolsep}{6pt}
\renewcommand{\arraystretch}{1.3}

\caption{Equatorial, polar, and volumetric deformations of each EOS with respect to the $T=0$ configuration for a purely poloidal magnetic field.}
\label{tab:pp}

\begin{tabular}{ccccccc}
\hline
Thermodynamic \\ conditions &
\begin{tabular}{c}
$R_{e}$ \\
(km)
\end{tabular} &
\begin{tabular}{c}
$R_{p}$ \\
(km)
\end{tabular} &
\begin{tabular}{c}
Volume \\
(km$^3$)
\end{tabular} &
\begin{tabular}{c}
Equatorial \\
deformation (\%)
\end{tabular} &
\begin{tabular}{c}
Polar \\
deformation (\%)
\end{tabular} &
\begin{tabular}{c}
Volumetric \\
deformation (\%)
\end{tabular} \\
\hline
$s_B=1,\;Y_l=0.4$     & 13.81 & 10.48 &  8626.74 & 20.70 & 16.76 & 71.01 \\
$s_B=2,\;Y_l=0.2$     & 14.67 &  9.93 &  9476.52 & 28.25 & 10.70 & 87.86 \\
$s_B=2,\;Y_{\nu_e}=0$ & 14.67 & 10.16 &  9621.82 & 28.25 & 13.19 & 90.74 \\
$T=0$                 & 11.44 &  8.97 &  5044.46 &  0.00 &  0.00 &  0.00 \\
\hline
\end{tabular}
\end{table*}

\begin{table*}[htbp]
\centering
\scriptsize
\setlength{\tabcolsep}{6pt}
\renewcommand{\arraystretch}{1.3}

\caption{Equatorial, polar and volumetric deformations of each EOS for a mixed-field configuration with $a=0.5$.}
\label{tab:mixed_a05}

\begin{tabular}{ccccccc}
\hline
Thermodynamic \\ conditions &
\begin{tabular}{c}
$R_{e}$ \\
(km)
\end{tabular} &
\begin{tabular}{c}
$R_{p}$ \\
(km)
\end{tabular} &
\begin{tabular}{c}
Volume \\
(km$^3$)
\end{tabular} &
\begin{tabular}{c}
Equatorial \\
deformation (\%)
\end{tabular} &
\begin{tabular}{c}
Polar \\
deformation (\%)
\end{tabular} &
\begin{tabular}{c}
Volumetric \\
deformation (\%)
\end{tabular} \\
\hline
$s_B=1,\;Y_l=0.4$     & 14.72 & 10.88 &  9524.03 & 28.33 & 15.71 & 85.19 \\
$s_B=2,\;Y_l=0.2$     & 15.01 & 11.07 & 10080.16 & 30.90 & 17.80 & 96.00 \\
$s_B=2,\;Y_{\nu_e}=0$ & 15.21 & 11.17 & 10384.76 & 32.62 & 18.85 & 101.93 \\
$T=0$                 & 11.47 &  9.40 &  5142.85 &  0.00 &  0.00 &   0.00 \\
\hline
\end{tabular}
\end{table*}

\begin{table*}[htbp]
\centering
\scriptsize
\setlength{\tabcolsep}{6pt}
\renewcommand{\arraystretch}{1.3}

\caption{Equatorial, polar and volumetric deformations of each EOS for a mixed-field configuration with $a=1.0$.}
\label{tab:mixed_a10}

\begin{tabular}{ccccccc}
\hline
Thermodynamic \\ conditions &
\begin{tabular}{c}
$R_{e}$ \\
(km)
\end{tabular} &
\begin{tabular}{c}
$R_{p}$ \\
(km)
\end{tabular} &
\begin{tabular}{c}
Volume \\
(km$^3$)
\end{tabular} &
\begin{tabular}{c}
Equatorial \\
deformation (\%)
\end{tabular} &
\begin{tabular}{c}
Polar \\
deformation (\%)
\end{tabular} &
\begin{tabular}{c}
Volumetric \\
deformation (\%)
\end{tabular} \\
\hline
$s_B=1,\;Y_l=0.4$     & 15.11 & 11.27 &  9806.35 & 22.31 & 26.52 & 83.22 \\
$s_B=2,\;Y_l=0.2$     & 16.19 & 11.07 & 10869.36 & 31.08 & 24.31 & 103.08 \\
$s_B=2,\;Y_{\nu_e}=0$ & 16.10 & 11.27 & 11004.61 & 30.28 & 26.52 & 105.60 \\
$T=0$                 & 12.35 &  8.91 &  5352.34 &  0.00 &  0.00 &   0.00 \\
\hline
\end{tabular}
\end{table*}

\begin{table*}[htbp]
\centering
\scriptsize
\setlength{\tabcolsep}{6pt}
\renewcommand{\arraystretch}{1.3}

\caption{Equatorial, polar and volumetric deformations of each EOS for a mixed-field configuration with $k_b=0.1$.}
\label{tab:mixed_kb01}

\begin{tabular}{ccccccc}
\hline
Thermodynamic \\ conditions &
\begin{tabular}{c}
$R_{e}$ \\
(km)
\end{tabular} &
\begin{tabular}{c}
$R_{p}$ \\
(km)
\end{tabular} &
\begin{tabular}{c}
Volume \\
(km$^3$)
\end{tabular} &
\begin{tabular}{c}
Equatorial \\
deformation (\%)
\end{tabular} &
\begin{tabular}{c}
Polar \\
deformation (\%)
\end{tabular} &
\begin{tabular}{c}
Volumetric \\
deformation (\%)
\end{tabular} \\
\hline
$s_B=1,\;Y_l=0.4$     & 13.24 & 12.45 & 9037.24 & 20.63 & 21.05 & 76.03 \\
$s_B=2,\;Y_l=0.2$     & 13.54 & 12.65 & 9530.47 & 23.32 & 22.97 & 85.63 \\
$s_B=2,\;Y_{\nu_e}=0$ & 13.63 & 12.75 & 9780.77 & 24.22 & 23.92 & 90.51 \\
$T=0$                 & 10.98 & 10.29 & 5134.00 &  0.00 &  0.00 &  0.00 \\
\hline
\end{tabular}
\end{table*}

\begin{table*}[htbp]
\centering
\scriptsize
\setlength{\tabcolsep}{6pt}
\renewcommand{\arraystretch}{1.3}

\caption{Equatorial, polar and volumetric deformations of each EOS for a mixed-field configuration with $k_b=0.04$.}
\label{tab:mixed_kb004}

\begin{tabular}{ccccccc}
\hline
Thermodynamic \\ conditions &
\begin{tabular}{c}
$R_{e}$ \\
(km)
\end{tabular} &
\begin{tabular}{c}
$R_{p}$ \\
(km)
\end{tabular} &
\begin{tabular}{c}
Volume \\
(km$^3$)
\end{tabular} &
\begin{tabular}{c}
Equatorial \\
deformation (\%)
\end{tabular} &
\begin{tabular}{c}
Polar \\
deformation (\%)
\end{tabular} &
\begin{tabular}{c}
Volumetric \\
deformation (\%)
\end{tabular} \\
\hline
$s_B=1,\;Y_l=0.4$     & 12.85 & 12.75 & 8964.95 & 20.28 & 20.47 & 75.62 \\
$s_B=2,\;Y_l=0.2$     & 13.14 & 12.95 & 9451.28 & 23.04 & 22.33 & 85.15 \\
$s_B=2,\;Y_{\nu_e}=0$ & 13.24 & 13.04 & 9646.34 & 23.96 & 23.26 & 88.97 \\
$T=0$                 & 10.68 & 10.58 & 5104.72 &  0.00 &  0.00 &  0.00 \\
\hline
\end{tabular}
\end{table*}

\newpage

% Bibliography

%% [A] Recommended: using JHEP.bst file
\bibliographystyle{JHEP}
\bibliography{biblio.bib}

@article{Braithwaite.x,
    author = {Braithwaite, Jonathan},
    title = {Axisymmetric magnetic fields in stars: relative strengths of poloidal and toroidal components},
    journal = {Monthly Notices of the Royal Astronomical Society},
    volume = {397},
    number = {2},
    pages = {763-774},
    year = {2009},
    month = {07},
    issn = {0035-8711},
    doi = {10.1111/j.1365-2966.2008.14034.x},
    url = {https://doi.org/10.1111/j.1365-2966.2008.14034.x},
    eprint = {https://academic.oup.com/mnras/article-pdf/397/2/763/2932157/mnras0397-0763.pdf},
}

@article{Kiuchi_2009,
doi = {10.1088/0004-637X/698/1/541},
url = {https://dx.doi.org/10.1088/0004-637X/698/1/541},
year = {2009},
month = {may},
publisher = {The American Astronomical Society},
volume = {698},
number = {1},
pages = {541},
author = {Kiuchi, Kenta and Kotake, Kei and Yoshida, Shijun},
title = {EQUILIBRIUM CONFIGURATIONS OF RELATIVISTIC STARS WITH PURELY TOROIDAL MAGNETIC FIELDS: EFFECTS OF REALISTIC EQUATIONS OF STATE},
journal = {The Astrophysical Journal},
}

@article{Frieben:2012dz,
    author = "Frieben, Joachim and Rezzolla, Luciano",
    title = "{Equilibrium models of relativistic stars with a toroidal magnetic field}",
    eprint = "1207.4035",
    archivePrefix = "arXiv",
    primaryClass = "gr-qc",
    doi = "10.1111/j.1365-2966.2012.22027.x",
    journal = "Mon. Not. Roy. Astron. Soc.",
    volume = "427",
    pages = "3406--3426",
    year = "2012"
}

@article{kiuchi2008.78.044045,
  title = {Relativistic stars with purely toroidal magnetic fields},
  author = {Kiuchi, Kenta and Yoshida, Shijun},
  journal = {Phys. Rev. D},
  volume = {78},
  issue = {4},
  pages = {044045},
  numpages = {21},
  year = {2008},
  month = {Aug},
  publisher = {American Physical Society},
  doi = {10.1103/PhysRevD.78.044045},
  url = {https://link.aps.org/doi/10.1103/PhysRevD.78.044045}
}

@article{Konno_2001,
  author = "Konno, K.",
    title = "{Moments of inertia of relativistic magnetized stars}",
    eprint = "gr-qc/0105015",
    archivePrefix = "arXiv",
    doi = "10.1051/0004-6361:20010556",
    journal = "Astron. Astrophys.",
    volume = "372",
    pages = "594",
    year = "2001"
}

@article{Tayler/mnras/161.4.365,
    author = {Tayler, R. J.},
    title = { The Adiabatic Stability of Stars Containing Magnetic Fields–I: T OROIDAL F IELDS},
    journal = {Monthly Notices of the Royal Astronomical Society},
    volume = {161},
    number = {4},
    pages = {365-380},
    year = {1973},
    month = {04},
    issn = {0035-8711},
    doi = {10.1093/mnras/161.4.365},
    url = {https://doi.org/10.1093/mnras/161.4.365},
    eprint = {https://academic.oup.com/mnras/article-pdf/161/4/365/8076545/mnras161-0365.pdf},
}

@article{Markey93/mnras/163.1.77,
    author = {Markey, P. and Tayler, R. J.},
    title = { The Adiabatic Stability of Stars Containing Magnetic Fields–II: POLOIDAL FIELDS},
    journal = {Monthly Notices of the Royal Astronomical Society},
    volume = {163},
    number = {1},
    pages = {77-91},
    year = {1973},
    month = {07},
    issn = {0035-8711},
    doi = {10.1093/mnras/163.1.77},
    url = {https://doi.org/10.1093/mnras/163.1.77},
    eprint = {https://academic.oup.com/mnras/article-pdf/163/1/77/8079493/mnras163-0077.pdf},
}

@article{ Braithwaite2006,
	 author = "Braithwaite, Jonathan and Spruit, H. C.",
    title = "{Evolution of the magnetic field in magnetars}",
    eprint = "astro-ph/0510287",
    archivePrefix = "arXiv",
    doi = "10.1051/0004-6361:20041981",
    journal = "Astron. Astrophys.",
    volume = "450",
    pages = "1097",
    year = "2006"
}

@article{ BraithwaiteId0,
	 author = "Braithwaite, Jonathan and Nordlund, A.",
    title = "{Stable magnetic fields in stellar interiors}",
    eprint = "astro-ph/0510316",
    archivePrefix = "arXiv",
    doi = "10.1051/0004-6361:20041980",
    journal = "Astron. Astrophys.",
    volume = "450",
    pages = "1077",
    year = "2006"
}

@article{Wright/mnras/162.4.339,
    author = {Wright, G. A. E.},
    title = {Pinch Instabilities in Magnetic Stars},
    journal = {Monthly Notices of the Royal Astronomical Society},
    volume = {162},
    number = {4},
    pages = {339-358},
    year = {1973},
    month = {06},
    issn = {0035-8711},
    doi = {10.1093/mnras/162.4.339},
    url = {https://doi.org/10.1093/mnras/162.4.339},
    eprint = {https://academic.oup.com/mnras/article-pdf/162/4/339/8073447/mnras162-0339.pdf},
}

@ARTICLE{pren...123..498P,
       author = {{Prendergast}, Kevin H.},
        title = "{The Equilibrium of a Self-Gravitating Incompressible Fluid Sphere with a Magnetic Field. I.}",
      journal = {\apj},
         year = 1956,
        month = may,
       volume = {123},
        pages = {498},
          doi = {10.1086/146186},
       adsurl = {https://ui.adsabs.harvard.edu/abs/1956ApJ...123..498P}
}

@article{Yaza85.044030,
  title = {Relativistic models of magnetars: Nonperturbative analytical approach},
  author = {Yazadjiev, Stoytcho S.},
  journal = {Phys. Rev. D},
  volume = {85},
  issue = {4},
  pages = {044030},
  numpages = {6},
  year = {2012},
  month = {Feb},
  publisher = {American Physical Society},
  doi = {10.1103/PhysRevD.85.044030},
  url = {https://link.aps.org/doi/10.1103/PhysRevD.85.044030}
}

@article{Bucciantini:2014zca,
    author = "Bucciantini, N. and Pili, A. G. and Del Zanna, L.",
    editor = "Pogorelov, N. V. and Audit, E. and Zank, G. P.",
    title = "{Solving the 3+1 GRMHD equations in the eXtended Conformally Flat Condition: the XNS code for magnetized neutron stars}",
    eprint = "1401.3101",
    archivePrefix = "arXiv",
    primaryClass = "astro-ph.HE",
    journal = "ASP Conf. Ser.",
    volume = "488",
    pages = "211--216",
    year = "2014"
}

@article{refId1,
	author = {{Bucciantini, N.} and {Del Zanna, L.}},
	title = {General relativistic magnetohydrodynamics in axisymmetric
          dynamical spacetimes: the X-ECHO code},
	DOI= "10.1051/0004-6361/201015945",
	url= "https://doi.org/10.1051/0004-6361/201015945",
	journal = {A\&A},
	year = 2011,
	volume = 528,
	pages = "A101",
}

@article{Cordero-Carrion:2008grk,
    author = "Cordero-Carrion, Isabel and Cerda-Duran, Pablo and Dimmelmeier, Harald and Jaramillo, Jose Luis and Novak, Jerome and Gourgoulhon, Eric",
    title = "{An Improved constrained scheme for the Einstein equations: An Approach to the uniqueness issue}",
    eprint = "0809.2325",
    archivePrefix = "arXiv",
    primaryClass = "gr-qc",
    doi = "10.1103/PhysRevD.79.024017",
    journal = "Phys. Rev. D",
    volume = "79",
    pages = "024017",
    year = "2009"
}

@article{Lander:2020bou,
    author = "Lander, S. K. and Haensel, P. and Haskell, B. and Zdunik, J. L. and Fortin, M.",
    title = "{Magnetic fields in late-stage proto-neutron stars}",
    eprint = "2007.14609",
    archivePrefix = "arXiv",
    primaryClass = "astro-ph.HE",
    doi = "10.1093/mnras/stab460",
    journal = "Mon. Not. Roy. Astron. Soc.",
    volume = "503",
    number = "1",
    pages = "875--895",
    year = "2021"
}

@article{Franzon:2016iai,
    author = "Franzon, B. and Dexheimer, V. and Schramm, S.",
    title = "{Internal composition of proto-neutron stars under strong magnetic fields}",
    eprint = "1606.04843",
    archivePrefix = "arXiv",
    primaryClass = "astro-ph.HE",
    doi = "10.1103/PhysRevD.94.044018",
    journal = "Phys. Rev. D",
    volume = "94",
    number = "4",
    pages = "044018",
    year = "2016"
}

@article{Mastrano:2013jaa,
    author = "Mastrano, Alpha and Lasky, Paul D. and Melatos, Andrew",
    title = "{Neutron star deformation due to multipolar magnetic fields}",
    eprint = "1306.4503",
    archivePrefix = "arXiv",
    primaryClass = "astro-ph.HE",
    doi = "10.1093/mnras/stt1131",
    journal = "Mon. Not. Roy. Astron. Soc.",
    volume = "434",
    pages = "1658",
    year = "2013"
}

@article{Mastrano:2015rfa,
    author = "Mastrano, Alpha and Suvorov, Arthur George and Melatos, Andrew",
    title = "{Neutron star deformation due to poloidal-toroidal magnetic fields of arbitrary multipole order: a new analytic approach}",
    eprint = "1501.01134",
    archivePrefix = "arXiv",
    primaryClass = "astro-ph.HE",
    doi = "10.1093/mnras/stu2671",
    journal = "Mon. Not. Roy. Astron. Soc.",
    volume = "447",
    pages = "3475",
    year = "2015"
}

@article{Colaiuda:2007br,
    author = "Colaiuda, A. and Ferrari, V. and Gualtieri, L. and Pons, J. A.",
    title = "{Relativistic models of magnetars: structure and deformations}",
    eprint = "0712.2162",
    archivePrefix = "arXiv",
    primaryClass = "astro-ph",
    doi = "10.1111/j.1365-2966.2008.12966.x",
    journal = "Mon. Not. Roy. Astron. Soc.",
    volume = "385",
    pages = "2080--2096",
    year = "2008"
}

@article{abramowski2011new,
  title={A new SNR with TeV shell-type morphology: HESS J1731-347},
  author={Abramowski, A and Acero, Fabio and Aharonian, F and Akhperjanian, AG and Anton, G and Balzer, Agn{\`e}s and Barnacka, Anna and De Almeida, U Barres and Becherini, Yvonne and Becker, J and others},
  journal={Astronomy \& Astrophysics},
  volume={531},
  pages={A81},
  year={2011},
  publisher={EDP Sciences}
}

@article{doroshenko2022strangely,
  title={A strangely light neutron star within a supernova remnant},
  author={Doroshenko, Victor and Suleimanov, Valery and P{\"u}hlhofer, Gerd and Santangelo, Andrea},
  journal={Nature Astronomy},
  volume={6},
  number={12},
  pages={1444--1451},
  year={2022},
  publisher={Nature Publishing Group UK London}
}

@article{Nakazato:2020ogl,
    author = "Nakazato, Ken'ichiro and Suzuki, Hideyuki",
    title = "{A New Approach to Mass and Radius of Neutron Stars with Supernova Neutrinos}",
    eprint = "2002.03300",
    archivePrefix = "arXiv",
    primaryClass = "astro-ph.HE",
    doi = "10.3847/1538-4357/ab7456",
    journal = "Astrophys. J.",
    volume = "891",
    pages = "156",
    year = "2020"
}

@article{Janka:2006fh,
    author = "Janka, Hans-Thomas and Langanke, K. and Marek, A. and Martinez-Pinedo, G. and Mueller, B.",
    title = "{Theory of Core-Collapse Supernovae}",
    eprint = "astro-ph/0612072",
    archivePrefix = "arXiv",
    doi = "10.1016/j.physrep.2007.02.002",
    journal = "Phys. Rept.",
    volume = "442",
    pages = "38--74",
    year = "2007"
}

@article{PhysRev.55.374,
  title = {On Massive Neutron Cores},
  author = {Oppenheimer, J. R. and Volkoff, G. M.},
  journal = {Phys. Rev.},
  volume = {55},
  issue = {4},
  pages = {374--381},
  numpages = {0},
  year = {1939},
  month = {Feb},
  publisher = {American Physical Society},
  doi = {10.1103/PhysRev.55.374},
  url = {https://link.aps.org/doi/10.1103/PhysRev.55.374}
}

@article{Lattimer_2001,
	doi = {10.1086/319702},
	url = {https://doi.org/10.1086%2F319702},
	year = {2001},
	publisher = {{IOP} Publishing},
	volume = {\textbf{550}},
	number = {1},
	pages = {426},
	author = {J. M. Lattimer and M. Prakash},
	journal = {The Astrophys. J.},
}

@article{Cardall_2001,
	doi = {10.1086/321370},
	url = {https://doi.org/10.1086/321370},
	year = 2001,
	month = {jun},
	publisher = {American Astronomical Society},
	volume = {554},
	number = {1},
	pages = {322--339},
	author = {Christian Y. Cardall and Madappa Prakash and James M. Lattimer},
	title = {Effects of Strong Magnetic Fields on Neutron Star Structure},
	journal = {The Astrophysical Journal},
}

@article{Prakash:1996xs,
    author = "Prakash, Madappa and Bombaci, Ignazio and Prakash, Manju and Ellis, Paul J. and Lattimer, James M. and Knorren, Roland",
    title = "{Composition and structure of protoneutron stars}",
    eprint = "nucl-th/9603042",
    archivePrefix = "arXiv",
    reportNumber = "SUNY-NTG-96-11, NUC-MINN-93-23-T",
    doi = "10.1016/S0370-1573(96)00023-3",
    journal = "Phys. Rept.",
    volume = "280",
    pages = "1--77",
    year = "1997"
}

@article{Pons:1998mm,
    author = "Pons, J. A. and Reddy, S. and Prakash, M. and Lattimer, J. M. and Miralles, J. A.",
    title = "{Evolution of protoneutron stars}",
    eprint = "astro-ph/9807040",
    archivePrefix = "arXiv",
    reportNumber = "SUNY-NTG-98-30",
    doi = "10.1086/306889",
    journal = "Astrophys. J.",
    volume = "513",
    pages = "780",
    year = "1999"
}

@article{Issifu:2023qoo,
    author = "Issifu, Adamu and da Silva, Franciele M. and Menezes, D\'ebora P.",
    title = "{Proto-strange quark stars from density-dependent quark mass model}",
    eprint = "2311.12511",
    archivePrefix = "arXiv",
    primaryClass = "nucl-th",
    doi = "10.1140/epjc/s10052-024-12828-0",
    journal = "Eur. Phys. J. C",
    volume = "84",
    number = "5",
    pages = "463",
    year = "2024"
}

@article{Roca-Maza:2011alv,
    author = "Roca-Maza, X. and Vinas, X. and Centelles, M. and Ring, P. and Schuck, P.",
    title = "{Relativistic mean field interaction with density dependent meson-nucleon vertices based on microscopical calculations}",
    eprint = "1110.2311",
    archivePrefix = "arXiv",
    primaryClass = "nucl-th",
    doi = "10.1103/PhysRevC.84.054309",
    journal = "Phys. Rev. C",
    volume = "84",
    pages = "054309",
    year = "2011",
    note = "[Erratum: Phys.Rev.C 93, 069905 (2016)]"
}

@article{Reed:2021nqk,
    author = "Reed, Brendan T. and Fattoyev, F. J. and Horowitz, C. J. and Piekarewicz, J.",
    title = "{Implications of PREX-2 on the Equation of State of Neutron-Rich Matter}",
    eprint = "2101.03193",
    archivePrefix = "arXiv",
    primaryClass = "nucl-th",
    doi = "10.1103/PhysRevLett.126.172503",
    journal = "Phys. Rev. Lett.",
    volume = "126",
    number = "17",
    pages = "172503",
    year = "2021"
}

@ARTICLE{riley2021,
      author = "Riley, Thomas E. and others",
    title = "{A NICER View of the Massive Pulsar PSR J0740+6620 Informed by Radio Timing and XMM-Newton Spectroscopy}",
    eprint = "2105.06980",
    archivePrefix = "arXiv",
    primaryClass = "astro-ph.HE",
    doi = "10.3847/2041-8213/ac0a81",
    journal = "Astrophys. J. Lett.",
    volume = "918",
    number = "2",
    pages = "L27",
    year = "2021"
}

@article{Camelio:2017nka,
    author = "Camelio, Giovanni and Lovato, Alessandro and Gualtieri, Leonardo and Benhar, Omar and Pons, Jos\'e A. and Ferrari, Valeria",
    title = "{Evolution of a proto-neutron star with a nuclear many-body equation of state: Neutrino luminosity and gravitational wave frequencies}",
    eprint = "1704.01923",
    archivePrefix = "arXiv",
    primaryClass = "astro-ph.HE",
    doi = "10.1103/PhysRevD.96.043015",
    journal = "Phys. Rev. D",
    volume = "96",
    number = "4",
    pages = "043015",
    year = "2017"
}

@ARTICLE{riley2019,
     author = "Riley, Thomas E. and others",
    title = "{A $NICER$ View of PSR J0030+0451: Millisecond Pulsar Parameter Estimation}",
    eprint = "1912.05702",
    archivePrefix = "arXiv",
    primaryClass = "astro-ph.HE",
    doi = "10.3847/2041-8213/ab481c",
    journal = "Astrophys. J. Lett.",
    volume = "887",
    number = "1",
    pages = "L21",
    year = "2019"
}

@article{Miller:2021qha,
    author = "Miller, M. C. and others",
    title = "{The Radius of PSR J0740+6620 from NICER and XMM-Newton Data}",
    eprint = "2105.06979",
    archivePrefix = "arXiv",
    primaryClass = "astro-ph.HE",
    doi = "10.3847/2041-8213/ac089b",
    journal = "Astrophys. J. Lett.",
    volume = "918",
    number = "2",
    pages = "L28",
    year = "2021"
}

@article{Issifu:2024fuw,
    author = "Issifu, Adamu and Menezes, D\'ebora P. and Rezaei, Zeinab and Frederico, Tobias",
    title = "{Proto-neutron stars with quark cores}",
    eprint = "2405.10386",
    archivePrefix = "arXiv",
    primaryClass = "nucl-th",
    doi = "10.1088/1475-7516/2025/01/024",
    journal = "JCAP",
    volume = "01",
    pages = "024",
    year = "2025"
}

@article{Issifu:2023qyi,
    author = "Issifu, Adamu and Marquez, Kauan D. and Pelicer, Mateus R. and Menezes, D\'ebora P.",
    title = "{Exotic baryons in hot neutron stars}",
    eprint = "2302.04364",
    archivePrefix = "arXiv",
    primaryClass = "nucl-th",
    doi = "10.1093/mnras/stad1198",
    journal = "Mon. Not. Roy. Astron. Soc.",
    volume = "522",
    number = "3",
    pages = "3263--3270",
    year = "2023"
}

@article{Prakash:2000jr,
    author = "Prakash, Madappa and Lattimer, James M. and Pons, Jose A. and Steiner, Andrew W. and Reddy, Sanjay",
    editor = "Blaschke, D. and Glendenning, N. K. and Sedrakian, A.",
    title = "{Evolution of a neutron star from its birth to old age}",
    eprint = "astro-ph/0012136",
    archivePrefix = "arXiv",
    journal = "Lect. Notes Phys.",
    volume = "578",
    pages = "364--423",
    year = "2001"
}

@article{Niksic:2002yp,
    author = "Niksic, T. and Vretenar, D. and Finelli, P. and Ring, P.",
    title = "{Relativistic Hartree-Bogolyubov model with density dependent meson nucleon couplings}",
    eprint = "nucl-th/0205009",
    archivePrefix = "arXiv",
    doi = "10.1103/PhysRevC.66.024306",
    journal = "Phys. Rev. C",
    volume = "66",
    pages = "024306",
    year = "2002"
}

@article{Typel:2018cap,
    author = "Typel, Stefan",
    editor = "Sedrakian, Armen",
    title = "{Relativistic Mean-Field Models with Different Parametrizations of Density Dependent Couplings}",
    doi = "10.3390/particles1010002",
    journal = "Particles",
    volume = "1",
    number = "1",
    pages = "3--22",
    year = "2018"
}

@article{Issifu:2024htq,
   author = "Issifu, Adamu and Thakur, Prashant and da Silva, Franciele M. and Marquez, Kau D. and Menezes, D\'ebora P. and Dutra, M. and Louren\c{c}o, O. and Frederico, Tobias",
    title = "{Supernova remnants with mirror dark matter and hyperons}",
    eprint = "2412.17946",
    archivePrefix = "arXiv",
    primaryClass = "hep-ph",
    doi = "10.1103/PhysRevD.111.083026",
    journal = "Phys. Rev. D",
    volume = "111",
    number = "8",
    pages = "083026",
    year = "2025"
}

@article{Menezes:2021jmw,
    author = "Menezes, D\'ebora Peres",
    title = "{A Neutron Star Is Born}",
    eprint = "2106.09515",
    archivePrefix = "arXiv",
    primaryClass = "astro-ph.HE",
    doi = "10.3390/universe7080267",
    journal = "Universe",
    volume = "7",
    number = "8",
    pages = "267",
    year = "2021"
}

@article{Serot:1997xg,
    author = "Serot, Brian D. and Walecka, John Dirk",
    title = "{Recent progress in quantum hadrodynamics}",
    eprint = "nucl-th/9701058",
    archivePrefix = "arXiv",
    reportNumber = "IU-NTC-96-17, JLAB-THY-97-48",
    doi = "10.1142/S0218301397000299",
    journal = "Int. J. Mod. Phys. E",
    volume = "6",
    pages = "515--631",
    year = "1997"
}

@article{Lattimer:2006xb,
    author = "Lattimer, James M. and Prakash, Maddapa",
    title = "{Neutron Star Observations: Prognosis for Equation of State Constraints}",
    eprint = "astro-ph/0612440",
    archivePrefix = "arXiv",
    doi = "10.1016/j.physrep.2007.02.003",
    journal = "Phys. Rept.",
    volume = "442",
    pages = "109--165",
    year = "2007"
}

@article{Lai:2000at,
    author = "Lai, Dong",
    title = "{Matter in strong magnetic fields}",
    eprint = "astro-ph/0009333",
    archivePrefix = "arXiv",
    doi = "10.1103/RevModPhys.73.629",
    journal = "Rev. Mod. Phys.",
    volume = "73",
    pages = "629--662",
    year = "2001"
}

@article{Pili:2014npa,
    author = "Pili, A.G. and Bucciantini, N. and Del Zanna, L.",
    title = "{Axisymmetric equilibrium models for magnetized neutron stars in General Relativity under the Conformally Flat Condition}",
    eprint = "1401.4308",
    archivePrefix = "arXiv",
    primaryClass = "astro-ph.HE",
    doi = "10.1093/mnras/stu215",
    journal = "Mon. Not. Roy. Astron. Soc.",
    volume = "439",
    pages = "3541--3563",
    year = "2014"
}

@ARTICLE{Heyl,
      author = "Heyl, Jeremy S. and Kulkarni, S. R.",
    title = "{How common are magnetars? the consequences of magnetic-field decay}",
    eprint = "astro-ph/9807306",
    archivePrefix = "arXiv",
    doi = "10.1086/311628",
    journal = "Astrophys. J. Lett.",
    volume = "506",
    pages = "L61",
    year = "1998"
}

@article{Bocquet:1995je,
    author = "Bocquet, M. and Bonazzola, S. and Gourgoulhon, E. and Novak, J.",
    title = "{Rotating neutron star models with magnetic field}",
    eprint = "gr-qc/9503044",
    archivePrefix = "arXiv",
    journal = "Astron. Astrophys.",
    volume = "301",
    pages = "757",
    year = "1995"
}

@article{Riley:2021pdl,
    author = "Riley, Thomas E. and others",
    title = "{A NICER View of the Massive Pulsar PSR J0740+6620 Informed by Radio Timing and XMM-Newton Spectroscopy}",
    eprint = "2105.06980",
    archivePrefix = "arXiv",
    primaryClass = "astro-ph.HE",
    doi = "10.3847/2041-8213/ac0a81",
    journal = "Astrophys. J. Lett.",
    volume = "918",
    number = "2",
    pages = "L27",
    year = "2021"
}

@article{Raduta:2020fdn,
    author = "Raduta, Adriana R. and Oertel, Micaela and Sedrakian, Armen",
    title = "{Proto-neutron stars with heavy baryons and universal relations}",
    eprint = "2008.00213",
    archivePrefix = "arXiv",
    primaryClass = "nucl-th",
    doi = "10.1093/mnras/staa2491",
    journal = "Mon. Not. Roy. Astron. Soc.",
    volume = "499",
    number = "1",
    pages = "914--931",
    year = "2020"
}

@article{Malfatti:2019tpg,
    author = "Malfatti, Germ\'an and Orsaria, Milva G. and Contrera, Gustavo A. and Weber, Fridolin and Ranea-Sandoval, Ignacio F.",
    title = "{Hot quark matter and (proto-) neutron stars}",
    eprint = "1907.06597",
    archivePrefix = "arXiv",
    primaryClass = "nucl-th",
    doi = "10.1103/PhysRevC.100.015803",
    journal = "Phys. Rev. C",
    volume = "100",
    number = "1",
    pages = "015803",
    year = "2019"
}

@article{Miller:2019cac,
    author = "Miller, M.C. and others",
    title = "{PSR J0030+0451 Mass and Radius from $NICER$ Data and Implications for the Properties of Neutron Star Matter}",
    eprint = "1912.05705",
    archivePrefix = "arXiv",
    primaryClass = "astro-ph.HE",
    doi = "10.3847/2041-8213/ab50c5",
    journal = "Astrophys. J. Lett.",
    volume = "887",
    number = "1",
    pages = "L24",
    year = "2019"
}

@article{Riley:2019yda,
    author = "Riley, Thomas E. and others",
    title = "{A $NICER$ View of PSR J0030+0451: Millisecond Pulsar Parameter Estimation}",
    eprint = "1912.05702",
    archivePrefix = "arXiv",
    primaryClass = "astro-ph.HE",
    doi = "10.3847/2041-8213/ab481c",
    journal = "Astrophys. J. Lett.",
    volume = "887",
    number = "1",
    pages = "L21",
    year = "2019"
}

@article{Oertel:2016bki,
    author = {Oertel, M. and Hempel, M. and Kl\"ahn, T. and Typel, S.},
    title = "{Equations of state for supernovae and compact stars}",
    doi = "10.1103/RevModPhys.89.015007",
    journal = "Rev. Mod. Phys.",
    volume = "89",
    number = "1",
    pages = "015007",
    year = "2017"
}

@ARTICLE{1991ApJ...383..745L,
       author = {{Lai}, Dong and {Shapiro}, Stuart L.},
        title = "{Cold Equation of State in a Strong Magnetic Field: Effects of Inverse beta-Decay}",
      journal = {\apj},
         year = 1991,
        month = dec,
       volume = {383},
        pages = {745},
          doi = {10.1086/170831}
}

@article{Ferrer,
  title = {Equation of state of a dense and magnetized fermion system},
  author = {Ferrer, Efrain J. and de la Incera, Vivian and Keith, Jason P. and Portillo, Israel and Springsteen, Paul L.},
  journal = {Phys. Rev. C},
  volume = {82},
  issue = {6},
  pages = {065802},
  numpages = {15},
  year = {2010},
  month = {Dec},
  publisher = {American Physical Society},
  doi = {10.1103/PhysRevC.82.065802},
  url = {https://link.aps.org/doi/10.1103/PhysRevC.82.065802}
}

@article{Kiuchi2008,
  title = {Relativistic stars with purely toroidal magnetic fields},
  author = {Kiuchi, Kenta and Yoshida, Shijun},
  journal = {Phys. Rev. D},
  volume = {78},
  issue = {4},
  pages = {044045},
  numpages = {21},
  year = {2008},
  month = {Aug},
  publisher = {American Physical Society},
  doi = {10.1103/PhysRevD.78.044045},
  url = {https://link.aps.org/doi/10.1103/PhysRevD.78.044045}
}

@article{ddme2,
  title = {New relativistic mean-field interaction with density-dependent meson-nucleon couplings},
  author = {Lalazissis, G. A. and Nik\ifmmode \check{s}\else \v{s}\fi{}i\ifmmode \acute{c}\else \'{c}\fi{}, T. and Vretenar, D. and Ring, P.},
  journal = {Phys. Rev. C},
  volume = {71},
  issue = {2},
  pages = {024312},
  numpages = {10},
  year = {2005},
  month = {Feb},
  publisher = {American Physical Society},
  doi = {10.1103/PhysRevC.71.024312},
  url = {https://link.aps.org/doi/10.1103/PhysRevC.71.024312}
}

@article{Choudhury:2024xbk,
doi = {10.3847/2041-8213/ad5a6f},
url = {https://dx.doi.org/10.3847/2041-8213/ad5a6f},
year = {2024},
month = {aug},
publisher = {The American Astronomical Society},
volume = {971},
number = {1},
pages = {L20},
author = {Devarshi Choudhury and others},
title = {A NICER View of the Nearest and Brightest Millisecond Pulsar: PSR J0437–4715},
journal = {The Astrophysical Journal Letters}
}

@article{Ciolfi_2009,
   title={Relativistic models of magnetars: the twisted torus magnetic field configuration},
   volume={397},
   ISSN={1365-2966},
   url={http://dx.doi.org/10.1111/j.1365-2966.2009.14990.x},
   DOI={10.1111/j.1365-2966.2009.14990.x},
   number={2},
   journal={Monthly Notices of the Royal Astronomical Society},
   publisher={Oxford University Press (OUP)},
   author={Ciolfi, R. and Ferrari, V. and Gualtieri, L. and Pons, J. A.},
   year={2009},
   month=aug, pages={913–924} }

@article{Goldreich,
author = {Goldreich, Peter and Reisenegger, Andreas},
year = {1992},
month = {09},
pages = {},
title = {Magnetic field decay in isolated neutron stars},
volume = {395},
journal = {ApJ},
doi = {10.1086/171646}
}

@article{Fischer:2009af,
    author = "Fischer, T. and Whitehouse, S. C. and Mezzacappa, A. and Thielemann, F. -K. and Liebendorfer, M.",
    title = "{Protoneutron star evolution and the neutrino driven wind in general relativistic neutrino radiation hydrodynamics simulations}",
    eprint = "0908.1871",
    archivePrefix = "arXiv",
    primaryClass = "astro-ph.HE",
    doi = "10.1051/0004-6361/200913106",
    journal = "Astron. Astrophys.",
    volume = "517",
    pages = "A80",
    year = "2010"
}

@article{Shao:2011nu,
    author = "Shao, Guo-yun",
    title = "{Evolution of proto-neutron stars with the hadron-quark phase transition}",
    eprint = "1109.4340",
    archivePrefix = "arXiv",
    primaryClass = "nucl-th",
    doi = "10.1016/j.physletb.2011.09.030",
    journal = "Phys. Lett. B",
    volume = "704",
    pages = "343--346",
    year = "2011"
}

@article{Burrows:1986me,
    author = "Burrows, Adam and Lattimer, James M.",
    title = "{The birth of neutron stars}",
    doi = "10.1086/164405",
    journal = "Astrophys. J.",
    volume = "307",
    pages = "178--196",
    year = "1986"
}

@ARTICLE{1989ApJ...342..958F,
       author = {{Fushiki}, I. and {Gudmundsson}, E.~H. and {Pethick}, C.~J.},
        title = "{Surface Structure of Neutron Stars with High Magnetic Fields}",
      journal = {\apj},
         year = 1989,
        month = jul,
       volume = {342},
        pages = {958},
          doi = {10.1086/167653},
       adsurl = {https://ui.adsabs.harvard.edu/abs/1989ApJ...342..958F}
}

@ARTICLE{1964ApJ...140.1309W,
       author = {{Woltjer}, L.},
        title = "{X-Rays and Type I Supernova Remnants.}",
      journal = {\apj},
         year = 1964,
        month = oct,
       volume = {140},
        pages = {1309-1313},
          doi = {10.1086/148028},
       adsurl = {https://ui.adsabs.harvard.edu/abs/1964ApJ...140.1309W}
}

@ARTICLE{1992ApJ...392L...9D,
  author  = {Duncan, Robert C. and Thompson, Christopher},
  title   = {Formation of Very Strongly Magnetized Neutron Stars: Implications for Gamma-Ray Bursts},
  journal = {The Astrophysical Journal Letters},
  year    = {1992},
  month   = {jun},
  volume  = {392},
  pages   = {L9},
  doi     = {10.1086/186413}
}

@article{Akiyama_2003,
doi = {10.1086/344135},
url = {https://dx.doi.org/10.1086/344135},
year = {2003},
month = {feb},
publisher = {},
volume = {584},
number = {2},
pages = {954},
author = {Akiyama, Shizuka and Wheeler, J. Craig and Meier, David L. and Lichtenstadt, Itamar},
title = {The Magnetorotational Instability in Core-Collapse Supernova Explosions},
journal = {The Astrophysical Journal}
}

@article{Obergaulinger_2009,
  author    = {Obergaulinger, M. and Cerdá-Durán, P. and Müller, E. and Aloy, M. A.},
  title     = {Semi-global simulations of the magneto-rotational instability in core collapse supernovae},
  journal   = {Astronomy \& Astrophysics},
  volume    = {498},
  number    = {1},
  pages     = {241--271},
  year      = {2009},
  month     = mar,
  publisher = {EDP Sciences},
  doi       = {10.1051/0004-6361/200811323},
  issn      = {1432-0746},
  url       = {http://dx.doi.org/10.1051/0004-6361/200811323}
}

@article{Sawai_2013,
doi = {10.1088/2041-8205/770/2/L19},
url = {https://dx.doi.org/10.1088/2041-8205/770/2/L19},
year = {2013},
month = {may},
publisher = {The American Astronomical Society},
volume = {770},
number = {2},
pages = {L19},
author = {Sawai, H. and Yamada, S. and Suzuki, H.}
}

@article{M_sta_2015,
   title={A large-scale dynamo and magnetoturbulence in rapidly rotating core-collapse supernovae},
   volume={528},
   ISSN={1476-4687},
   url={http://dx.doi.org/10.1038/nature15755},
   DOI={10.1038/nature15755},
   number={7582},
   journal={Nature},
   publisher={Springer Science and Business Media LLC},
   author={Mösta, Philipp and Ott, Christian D. and Radice, David and Roberts, Luke F. and Schnetter, Erik and Haas, Roland},
   year={2015},
   month=nov, pages={376–379} }

@article{Lattimer_2004,
   title={The Physics of Neutron Stars},
   volume={304},
   ISSN={1095-9203},
   url={http://dx.doi.org/10.1126/science.1090720},
   DOI={10.1126/science.1090720},
   number={5670},
   journal={Science},
   publisher={American Association for the Advancement of Science (AAAS)},
   author={Lattimer, J. M. and Prakash, M.},
   year={2004},
   month=apr, pages={536–542} }

@article{Cutler_2002,
   title={Gravitational waves from neutron stars with large toroidal<mml:math xmlns:mml="http://www.w3.org/1998/Math/MathML" display="inline"><mml:mi>B</mml:mi></mml:math>fields},
   volume={66},
   ISSN={1089-4918},
   url={http://dx.doi.org/10.1103/PhysRevD.66.084025},
   DOI={10.1103/physrevd.66.084025},
   number={8},
   journal={Physical Review D},
   publisher={American Physical Society (APS)},
   author={Cutler, Curt},
   year={2002},
   month=oct }

@ARTICLE{2011CQGra..28k4014G,
       author = {{Gualtieri}, L. and {Ciolfi}, R. and {Ferrari}, V.},
        title = "{Structure, deformations and gravitational wave emission of magnetars}",
      journal = {Classical and Quantum Gravity},
         year = 2011,
        month = jun,
       volume = {28},
       number = {11},
          eid = {114014},
        pages = {114014},
          doi = {10.1088/0264-9381/28/11/114014},
archivePrefix = {arXiv},
       eprint = {1011.2778},
 primaryClass = {astro-ph.SR},
       adsurl = {https://ui.adsabs.harvard.edu/abs/2011CQGra..28k4014G}
}

@ARTICLE{2007Ap&SS.308..119D,
       author       = {Dall'Osso, S. and Stella, L.},
  title        = {Newborn magnetars as sources of gravitational radiation: Constraints from high energy observations of magnetar candidates},
  journal      = {Astrophysics and Space Science},
  year         = {2007},
  month        = {April},
  volume       = {308},
  number       = {1--4},
  pages        = {119--124},
  doi          = {10.1007/s10509-007-9323-0},
  archivePrefix= {arXiv},
  eprint       = {astro-ph/0702075},
  primaryClass = {astro-ph},
  adsurl       = {https://ui.adsabs.harvard.edu/abs/2007Ap&SS.308..119D}
}

@article{Mastrano,
    author = {Mastrano, A. and Melatos, A. and Reisenegger, A. and Akgün, T.},
    title = {Gravitational wave emission from a magnetically deformed non-barotropic neutron star},
    journal = {Monthly Notices of the Royal Astronomical Society},
    volume = {417},
    number = {3},
    pages = {2288-2299},
    year = {2011},
    month = {10},
    issn = {0035-8711},
    doi = {10.1111/j.1365-2966.2011.19410.x},
    url = {https://doi.org/10.1111/j.1365-2966.2011.19410.x},
    eprint = {https://academic.oup.com/mnras/article-pdf/417/3/2288/3826519/mnras0417-2288.pdf},
}

@ARTICLE{Bethe,
       author = {{Bethe}, H.~A. and {Wilson}, J.~R.},
        title = "{Revival of a stalled supernova shock by neutrino heating}",
      journal = {\apj},
         year = 1985,
        month = aug,
       volume = {295},
        pages = {14-23},
          doi = {10.1086/163343},
       adsurl = {https://ui.adsabs.harvard.edu/abs/1985ApJ...295...14B}
}

@article{Mazurek1975,
    author = {{Mazurek}, T.~J.},
    title = "{Chemical potential effects on neutrino diffusion in supernovae}",
    journal = {Astrophysics and Space Science},
    volume = {35},
    pages = {117-137},
    year = {1975},
    doi = {10.1007/BF00639985},
    adsurl = {https://ui.adsabs.harvard.edu/abs/1975Ap&SS..35..117M}
}

@ARTICLE{1995A&A...296..145K,
      author       = {Keil, W. and Janka, H.-T.},
  title        = {Hadronic phase transitions at supranuclear densities and the delayed collapse of newly formed neutron stars},
  journal      = {Astronomy and Astrophysics},
  year         = {1995},
  volume       = {296},
  pages        = {145--+},
  month        = {April},
  adsurl       = {https://ui.adsabs.harvard.edu/abs/1995A&A...296..145K}
}

@article{PhysRevLett.104.251101,
  title = {Neutrino Signal of Electron-Capture Supernovae from Core Collapse to Cooling},
  author = {H\"udepohl, L. and M\"uller, B. and Janka, H.-T. and Marek, A. and Raffelt, G. G.},
  journal = {Phys. Rev. Lett.},
  volume = {104},
  issue = {25},
  pages = {251101},
  numpages = {4},
  year = {2010},
  month = {Jun},
  publisher = {American Physical Society},
  doi = {10.1103/PhysRevLett.104.251101},
  url = {https://link.aps.org/doi/10.1103/PhysRevLett.104.251101}
}

@inbook{Roberts_2017,
   title={Neutrino Signatures from Young Neutron Stars},
   ISBN={9783319218465},
   url={http://dx.doi.org/10.1007/978-3-319-21846-5_5},
   DOI={10.1007/978-3-319-21846-5_5},
   booktitle={Handbook of Supernovae},
   publisher={Springer International Publishing},
   author={Roberts, Luke F. and Reddy, Sanjay},
   year={2017},
   pages={1605–1635} }

@inproceedings{Spruit_2008,
   title={Origin of neutron star magnetic fields},
   volume={983},
   ISSN={0094-243X},
   url={http://dx.doi.org/10.1063/1.2900262},
   DOI={10.1063/1.2900262},
   booktitle={AIP Conference Proceedings},
   publisher={AIP},
   author={Spruit, H. C. and Bassa, C. and Wang, Z. and Cumming, A. and Kaspi, V. M.},
   year={2008},
   pages={391–398} }

@article{Ferrario_2015,
   title={Magnetic Field Generation in Stars},
   volume={191},
   ISSN={1572-9672},
   url={http://dx.doi.org/10.1007/s11214-015-0138-y},
   DOI={10.1007/s11214-015-0138-y},
   number={1–4},
   journal={Space Science Reviews},
   publisher={Springer Science and Business Media LLC},
   author={Ferrario, Lilia and Melatos, Andrew and Zrake, Jonathan},
   year={2015},
   month=mar, pages={77–109} }

@article{Mereghetti_2008,
   title={The strongest cosmic magnets: soft gamma-ray repeaters and anomalous X-ray pulsars},
   volume={15},
   ISSN={1432-0754},
   url={http://dx.doi.org/10.1007/s00159-008-0011-z},
   DOI={10.1007/s00159-008-0011-z},
   number={4},
   journal={The Astronomy and Astrophysics Review},
   publisher={Springer Science and Business Media LLC},
   author={Mereghetti, Sandro},
   year={2008},
   month=jul, pages={225–287} }

@ARTICLE{Kaspi,
       author       = {Kaspi, Victoria M. and Beloborodov, Andrei M.},
  title        = {Magnetars},
  journal      = {Annual Review of Astronomy and Astrophysics},
  year         = {2017},
  volume       = {55},
  number       = {1},
  pages        = {261--301},
  month        = {August},
  doi          = {10.1146/annurev-astro-081915-023329},
  archivePrefix= {arXiv},
  eprint       = {1703.00068},
  primaryClass = {astro-ph.HE},
  adsurl       = {https://ui.adsabs.harvard.edu/abs/2017ARA&A..55..261K}
}

@article{Chamel_2008,
   title={Physics of Neutron Star Crusts},
   volume={11},
   ISSN={1433-8351},
   url={http://dx.doi.org/10.12942/lrr-2008-10},
   DOI={10.12942/lrr-2008-10},
   number={1},
   journal={Living Reviews in Relativity},
   publisher={Springer Science and Business Media LLC},
   author={Chamel, Nicolas and Haensel, Pawel},
   year={2008},
   month=dec }

@article{RevModPhys.80.1455,
  title = {Color superconductivity in dense quark matter},
  author = {Alford, Mark G. and Schmitt, Andreas and Rajagopal, Krishna and Sch\"afer, Thomas},
  journal = {Rev. Mod. Phys.},
  volume = {80},
  issue = {4},
  pages = {1455--1515},
  numpages = {0},
  year = {2008},
  month = {Nov},
  publisher = {American Physical Society},
  doi = {10.1103/RevModPhys.80.1455},
  url = {https://link.aps.org/doi/10.1103/RevModPhys.80.1455}
}

@misc{bonazzola1996gravitationalwavesneutronstars,
      title={Gravitational waves from neutron stars}, 
      author={S. Bonazzola and E. Gourgoulhon},
      year={1996},
      eprint={astro-ph/9605187},
      archivePrefix={arXiv},
      primaryClass={astro-ph},
      url={https://arxiv.org/abs/astro-ph/9605187}, 
}

@ARTICLE{1953ApJ...118..116C,
       author = {{Chandrasekhar}, S. and {Fermi}, E.},
        title = "{Problems of Gravitational Stability in the Presence of a Magnetic Field.}",
      journal = {\apj},
         year = 1953,
        month = jul,
       volume = {118},
        pages = {116},
          doi = {10.1086/145732},
       adsurl = {https://ui.adsabs.harvard.edu/abs/1953ApJ...118..116C}
}

@ARTICLE{2022NatAs...6.1444D,
       author = {{Doroshenko}, Victor and {Suleimanov}, Valery and {P{\"u}hlhofer}, Gerd and {Santangelo}, Andrea},
        title = "{A strangely light neutron star within a supernova remnant}",
      journal = {Nature Astronomy},
         year = 2022,
        month = dec,
       volume = {6},
        pages = {1444-1451},
          doi = {10.1038/s41550-022-01800-1},
       adsurl = {https://ui.adsabs.harvard.edu/abs/2022NatAs...6.1444D}
}

@article{PhysRevLett.79.2176,
  title = {Quantizing Magnetic Field and Quark-Hadron Phase Transition in a Neutron Star},
  author = {Bandyopadhyay, Debades and Chakrabarty, Somenath and Pal, Subrata},
  journal = {Phys. Rev. Lett.},
  volume = {79},
  issue = {12},
  pages = {2176--2179},
  numpages = {0},
  year = {1997},
  month = {Sep},
  publisher = {American Physical Society},
  doi = {10.1103/PhysRevLett.79.2176},
  url = {https://link.aps.org/doi/10.1103/PhysRevLett.79.2176}
}

@article{PhysRevD.54.1306,
  title = {Quark matter in a strong magnetic field},
  author = {Chakrabarty, Somenath},
  journal = {Phys. Rev. D},
  volume = {54},
  issue = {2},
  pages = {1306--1316},
  numpages = {0},
  year = {1996},
  month = {Jul},
  publisher = {American Physical Society},
  doi = {10.1103/PhysRevD.54.1306},
  url = {https://link.aps.org/doi/10.1103/PhysRevD.54.1306}
}

@article{P_rez_Azor_n_2006,
   title={Anisotropic thermal emission from magnetized neutron stars},
   volume={451},
   ISSN={1432-0746},
   url={http://dx.doi.org/10.1051/0004-6361:20054403},
   DOI={10.1051/0004-6361:20054403},
   number={3},
   journal={Astronomy \&; Astrophysics},
   publisher={EDP Sciences},
   author={Pérez-Azorín, J. F. and Miralles, J. A. and Pons, J. A.},
   year={2006},
   month=may, pages={1009–1024} }

@article{Luo:2024qmq,
    author = "Luo, Yudong and Zha, Shuai and Kajino, Toshitaka",
    title = "{Strong magnetic field inside degenerate relativistic plasma and the impacts on the neutrino transport in Core-Collapse Supernovae}",
    eprint = "2405.11555",
    archivePrefix = "arXiv",
    primaryClass = "astro-ph.HE",
    month = "5",
    year = "2024"
}

@article{PhysRevC.107.065804,
  title = {Neutrino mean free path for proto--neutron star matter within the Skyrme model},
  author = {Bauer, E. and Torres Pati\~no, J. and Benvenuto, O. G.},
  journal = {Phys. Rev. C},
  volume = {107},
  issue = {6},
  pages = {065804},
  numpages = {21},
  year = {2023},
  month = {Jun},
  publisher = {American Physical Society},
  doi = {10.1103/PhysRevC.107.065804},
  url = {https://link.aps.org/doi/10.1103/PhysRevC.107.065804}
}

@article{Sur:2021awe,
    author = "Sur, Ankan and Cook, William and Radice, David and Haskell, Brynmor and Bernuzzi, Sebastiano",
    title = "{Long-term general relativistic magnetohydrodynamics simulations of magnetic field in isolated neutron stars}",
    eprint = "2108.11858",
    archivePrefix = "arXiv",
    primaryClass = "astro-ph.HE",
    doi = "10.1093/mnras/stac353",
    journal = "Mon. Not. Roy. Astron. Soc.",
    volume = "511",
    number = "3",
    pages = "3983--3993",
    year = "2022"
}

@article{Cheong:2024stz,
    author = "Cheong, Patrick Chi-Kit and Tsokaros, Antonios and Ruiz, Milton and Venturi, Fabrizio and Chan, Juno Chun Lung and Yip, Anson Ka Long and Uryu, Koji",
    title = "{General-relativistic resistive-magnetohydrodynamics simulations of self-consistent magnetized rotating neutron stars}",
    eprint = "2409.10508",
    archivePrefix = "arXiv",
    primaryClass = "astro-ph.HE",
    doi = "10.1103/PhysRevD.111.063030",
    journal = "Phys. Rev. D",
    volume = "111",
    number = "6",
    pages = "063030",
    year = "2025"
}

@article{Das:2022oyn,
    author = "Das, Pushpita and Porth, Oliver and Watts, Anna L.",
    title = "{GRMHD simulations of accreting neutron stars with non-dipole fields}",
    eprint = "2204.00249",
    archivePrefix = "arXiv",
    primaryClass = "astro-ph.HE",
    doi = "10.1093/mnras/stac1817",
    journal = "Mon. Not. Roy. Astron. Soc.",
    volume = "515",
    number = "3",
    pages = "3144--3161",
    year = "2022"
}

@article{Chatterjee:2021wsr,
    author = "Chatterjee, Debarati and Novak, J{\'e}r{\^o}me and Oertel, Micaela",
    title = "{Structure of ultra-magnetised neutron stars}",
    eprint = "2108.13733",
    archivePrefix = "arXiv",
    primaryClass = "nucl-th",
    doi = "10.1140/epja/s10050-021-00525-5",
    journal = "Eur. Phys. J. A",
    volume = "57",
    number = "8",
    pages = "249",
    year = "2021"
}

@article{Chatterjee:2014qsa,
    author = "Chatterjee, Debarati and Elghozi, Thomas and Novak, Jerome and Oertel, Micaela",
    title = "{Consistent neutron star models with magnetic field dependent equations of state}",
    eprint = "1410.6332",
    archivePrefix = "arXiv",
    primaryClass = "astro-ph.HE",
    doi = "10.1093/mnras/stu2706",
    journal = "Mon. Not. Roy. Astron. Soc.",
    volume = "447",
    pages = "3785",
    year = "2015"
}

@article{Rather:2022bmm,
    author = "Rather, Ishfaq A. and Rather, Asloob A. and Dexheimer, V. and Lopes, Il{\'\i}dio and Usmani, A. A. and Patra, S. K.",
    title = "{Magnetic-field Induced Deformation in Hybrid Stars}",
    eprint = "2209.06016",
    archivePrefix = "arXiv",
    primaryClass = "nucl-th",
    doi = "10.3847/1538-4357/aca85c",
    journal = "Astrophys. J.",
    volume = "943",
    number = "1",
    pages = "52",
    year = "2023"
}

@article{Cardall:2000bs,
    author = "Cardall, Christian Y. and Prakash, Madappa and Lattimer, James M.",
    title = "{Effects of strong magnetic fields on neutron star structure}",
    eprint = "astro-ph/0011148",
    archivePrefix = "arXiv",
    doi = "10.1086/321370",
    journal = "Astrophys. J.",
    volume = "554",
    pages = "322--339",
    year = "2001"
}

@article{Kiuchi:2008ss,
    author = "Kiuchi, Kenta and Shibata, Masaru and Yoshida, Shijun",
    title = "{Evolution of neutron stars with toroidal magnetic fields: Axisymmetric simulation in full general relativity}",
    eprint = "0805.2712",
    archivePrefix = "arXiv",
    primaryClass = "astro-ph",
    doi = "10.1103/PhysRevD.78.024029",
    journal = "Phys. Rev. D",
    volume = "78",
    pages = "024029",
    year = "2008"
}

@article{Ioka:2003nh,
    author = "Ioka, Kunihito and Sasaki, Misao",
    title = "{Relativistic stars with poloidal and toroidal magnetic fields and meridional flow}",
    eprint = "astro-ph/0305352",
    archivePrefix = "arXiv",
    doi = "10.1086/379650",
    journal = "Astrophys. J.",
    volume = "600",
    pages = "296--316",
    year = "2004"
}

@article{Scurto:2024xxo,
    author = "Scurto, Luigi and Carvalho, Val{\'e}ria and Pais, Helena and Provid{\^e}ncia, Constan{\c{c}}a",
    title = "{Assessing the joint effect of temperature and magnetic field on the neutron star equation~of state}",
    eprint = "2407.03113",
    archivePrefix = "arXiv",
    primaryClass = "nucl-th",
    doi = "10.1103/PhysRevC.110.045805",
    journal = "Phys. Rev. C",
    volume = "110",
    number = "4",
    pages = "045805",
    year = "2024"
}

@article{Malik:2018blm,
    author = "Malik, Tuhin and Sen, Debashree and Jha, T. K. and Mishra, Hiranmaya",
    title = "{Impact of Magnetic field on neutron star properties}",
    eprint = "1811.06258",
    archivePrefix = "arXiv",
    primaryClass = "nucl-th",
    month = "11",
    year = "2018"
}

%% or
%% [B] Manual formatting (see below)
%% (i) We suggest to always provide author, title and journal data or doi:
%% in short all the informations that clearly identify a document.
%% (ii) please avoid comments such as "For a review'', "For some examples",
%% "and references therein" or move them in the text. In general, please leave only references in the bibliography and move all
%% accessory text in footnotes.
%% (iii) Also, please have only one work for each \bibitem.

\end{document}